\newif\ifanon

\newif\ifdraft

\newif\ifieee

\newif\ifccs

\newif\iflipics

\newif\iflncs
\lncstrue

\newif\ifsubmission
\submissiontrue

\newif\ifshort
\shorttrue

\PassOptionsToPackage{hyphens,spaces,obeyspaces}{url}

\iflipics
\documentclass[a4paper,UKenglish,cleveref, autoref, thm-restate]{format/lipics-v2021}

\ifanon

\author{Anonymous}{Anonymous}{}{}{}
\authorrunning{Anonymous}
\Copyright{Anonymous}

\else

\author{Dimitris Karakostas}{University of Edinburgh, United Kingdom}{dkarakos@inf.ed.ac.uk}{}{}
\author{Aggelos Kiayias}{University of Edinburgh, United Kingdom \and IOG, United Kingdom}{akiayias@inf.ed.ac.uk}{}{}
\author{Christina Ovezik}{University of Edinburgh, United
Kingdom}{c.ovezik@ed.ac.uk}{}{}

\authorrunning{D. Karakostas and A. Kiayias and C. Ovezik} 
\Copyright{Dimitris Karakostas and Aggelos Kiayias and Christina Ovezik} 

\fi



\category{} 

\relatedversion{} 

\nolinenumbers 

\EventEditors{John Q. Open and Joan R. Access}
\EventNoEds{2}
\EventLongTitle{42nd Conference on Very Important Topics (CVIT 2016)}
\EventShortTitle{CVIT 2016}
\EventAcronym{CVIT}
\EventYear{2016}
\EventDate{December 24--27, 2016}
\EventLocation{Little Whinging, United Kingdom}
\EventLogo{}
\SeriesVolume{42}
\ArticleNo{23}

\else

\ifccs
    \ifanon
        \documentclass[sigconf,anonymous]{acmart}
    \else
        \documentclass[sigconf]{acmart}
    \fi

    \AtBeginDocument{%
      \providecommand\BibTeX{{%
        \normalfont B\kern-0.5em{\scshape i\kern-0.25em b}\kern-0.8em\TeX}}}

    \setcopyright{acmcopyright}
    \copyrightyear{2022}
    \acmYear{2022}
    \acmDOI{XXXXXXX.XXXXXXX}

    \acmConference[CCS '22]{ACM Conference on Computer and Communications Security}{November 7-11, 2022}{Los Angeles, U.S.A.}
    \acmPrice{15.00}
    \acmISBN{978-1-4503-XXXX-X/18/06}

\else
    \ifieee
        \documentclass[conference,compsoc]{format/IEEEtran}
        \IEEEoverridecommandlockouts

        \usepackage{amsthm}
        \newtheorem{theorem}{Theorem}
        \newtheorem{lemma}{Lemma}
        \newtheorem{definition}{Definition}
        \newtheorem{proposition}{Proposition}

        \usepackage[colorlinks=true,citecolor=black,linkcolor=black,urlcolor=black,bookmarksopen=true]{hyperref}
    \else
        \iflncs
            \documentclass[runningheads]{format/llncs}

            \makeatother
            \usepackage[hidelinks, bookmarksopen=true]{hyperref}
            \usepackage{bookmark}
        \else
            \documentclass[10pt,a4paper]{article}
            \usepackage[utf8]{inputenc}

            \usepackage{amsthm}

            \newtheorem{definition}{Definition}

            \makeatother
            \usepackage[hidelinks, bookmarksopen=true]{hyperref}
            \usepackage{bookmark}

            \usepackage{geometry}
        \fi
    \fi

    \usepackage{amsmath}
    \usepackage{amssymb}
    \usepackage{xcolor}
    \usepackage{multirow}

\fi

\fi

\usepackage{xifthen}
\usepackage{paralist}
\usepackage{array,multirow,graphicx}
\usepackage{booktabs}
\usepackage{arydshln}

\newcommand{\discuss}[1]{{\color{red} #1}}

\newcommand{\ignore}[1]{}

\newcommand{\eg}{e.g., }
\newcommand{\ie}{i.e., }
\newcommand{\wrt}{w.r.t. }
\newcommand{\etal}{\textit{et al.}}
\newcommand{\etc}{\textit{etc.}}
\iflipics\else

\fi

\usepackage{tikz}

\ifieee\else
    
\begin{document}
\fi

\title{SoK: A Stratified Approach to Blockchain Decentralization 
    \ifdraft
    \\
    \discuss{Internal draft - Please do not distribute}
    \fi
}

\iflipics

\else

\ifccs
	\author{ Christina Ovezik }
	\affiliation{ 
		\institution{University of Edinburgh}
	}
	\email{ c.ovezik@ed.ac.uk }
    \author{ Dimitris Karakostas }
    \affiliation{ 
    \institution{University of Edinburgh}
    }
    \email{ dkarakos@inf.ed.ac.uk }

    \author{ Aggelos Kiayias }
    \affiliation{ 
    \institution{ U. of Edinburgh and IOG }
    }
    \email{ akiayias@inf.ed.ac.uk }
    
\else
    \ifieee
        \ifanon
            \author{
                \IEEEauthorblockN{Anonymous}
                \IEEEauthorblockA{}
            }
        \else
            \author{
            	\IEEEauthorblockN{Christina Ovezik}
            	\IEEEauthorblockA{University of Edinburgh \\ 
            		c.ovezik@ed.ac.uk}
            	\and
                \IEEEauthorblockN{Dimitris Karakostas}
                \IEEEauthorblockA{University of Edinburgh \\ dkarakos@inf.ed.ac.uk}
                \and
                \IEEEauthorblockN{Aggelos Kiayias}
                \IEEEauthorblockA{U. of Edinburgh and IOG \\ 
                akiayias@inf.ed.ac.uk}
            }
        \fi
    \else
        \iflncs
            \ifanon
                \institute{}
                \author{Anonymous}
            \else
                \author{
                	Christina Ovezik\inst{1}
                	\thanks{The order of the authors follows the Blockchain 
                	Technology Laboratory's Author Ordering Policy (           	
                	\url{https://www.ed.ac.uk/informatics/blockchain/btl-papers/aop}).}
                	\and
                    Dimitris Karakostas\inst{1}
                    \and
                    Aggelos Kiayias\inst{1,2}
                }
                \institute{
                    University of Edinburgh \and
                    IOG \\
                \email{\{c.ovezik,dkarakos,akiayias\}@ed.ac.uk}
            }
            \fi
        \else
        \iflipics
        \else
        \author{
            Christina Ovezik \\ U. of Edinburgh \\ c.ovezik@ed.ac.uk
            \and 
            Dimitris Karakostas \\ U. of Edinburgh \\ dkarakos@inf.ed.ac.uk
            \and
            Aggelos Kiayias \\ U. of Edinburgh and IOG \\ akiayias@inf.ed.ac.uk
           
        }
        \fi
        \fi
    \fi
\fi

\fi

\ifieee
    \begin{document}
    \maketitle
    \thispagestyle{plain}
    \pagestyle{plain}
\fi

\ifccs\else
\ifieee\else
    \maketitle
\fi
\fi

\begin{abstract}
Decentralization has been touted as the principal 
security advantage which propelled  blockchain
systems at the forefront of developments in the financial technology
space. 
Its exact semantics nevertheless remain highly contested and ambiguous,
with proponents and critics disagreeing widely on the level of decentralization
offered by existing systems.
To address this, we put forth a systematization of the current landscape with respect to decentralization and we derive a methodology that can help direct future research towards defining and measuring decentralization.
Our approach dissects blockchain systems into multiple layers, or strata, each 
possibly encapsulating multiple categories, and it enables a unified method for 
measuring decentralization in each one.
Our layers are 
    \begin{inparaenum}[(1)]
        \item hardware,
        \item software,
        \item network,
        \item consensus,
        \item economics (``tokenomics''),
        \item client API,
        \item governance, and
        \item geography.
    \end{inparaenum}
Armed with this stratification, we examine for each layer which pertinent 
properties of distributed ledgers (safety, liveness, privacy, stability) can be 
at risk due to centralization and in what way.
%
%
%
We also introduce a practical test, the ``Minimum Decentralization Test''  
which can provide quick insights about the decentralization state of a 
blockchain system. 
To demonstrate how our stratified methodology can be used in practice, we apply 
it fully (layer by layer) to Bitcoin, and we provide examples of systems which 
comprise one or more ``problematic'' layers that cause them to fail the MDT. 
Our work highlights the challenges in measuring and achieving decentralization, 
\ifsubmission\else
points to the degree of (de)centralization of various existing systems,
where such assessment can be made from presently available public information, 
\fi
and suggests various potential directions where future research is needed.
\end{abstract}

\ifsubmission
\ifieee
\fi
\fi

\ifccs 


\maketitle
\fi

\section{Introduction}\label{sec:introduction}

Bitcoin~\cite{nakamoto2008bitcoin}, the first blockchain-based distributed
ledger,\footnote{For the rest of this work we use the terms ``blockchain'' and
``distributed ledger'' interchangeably, even though strictly speaking, the
latter describes an objective while the former is a means to it.} put forth a
new paradigm, that inspired numerous systems to enhance and expand its model
and thousands of applications to be built on them. Alongside, a research
discipline emerged across cryptography, distributed systems, game theory and
economics, to analyze the properties and capabilities of this paradigm-shifting
protocol. 
 
Bitcoin's arguably most important contribution was offering a solution
to the consensus problem~\cite{lamport1982byzantine,pease1980reaching} in an
open setting. Contrary to classic protocols, cf.~\cite{RSA:GarKia20},
Bitcoin participants are not known a priori; instead, the system only assumes a
peer-to-peer (P2P) synchronous network and a public setup.\footnote{Bitcoin
uses the following newspaper headline as the common
setup string: ``The Times 03/Jan/2009 Chancellor on brink of second bailout for
banks.''} 
%
Bitcoin's core security argument is that, if a majority of computational
power acts honestly, the protocol solves the consensus problem and implements a
distributed ledger, as shown formally 
in~\cite{EC:GarKiaLeo15,EC:PasSeeShe17,C:GarKiaLeo17}.
This, in conjunction with the  premise that computational power is widely 
distributed over the network participants, gives rise to the 
``security via decentralization'' proposition: 
the system has no single point of failure, as any network participant 
is individually too weak to influence the properties of the protocol,
no matter how they behave. 
\ifshort\else
Consequently, ensuring that an adversary is not able to corrupt enough parties to 
control a majority (or that such a majority is not economically incentivized to collude and deviate from the protocol) is the fundamental issue that blockchain systems should 
establish for practical security. 
\fi
Intuitively, a high degree of decentralization suggests that the 
trust for safe system operation is spread 
across the largest possible set of parties. 

The appeal of this narrative, and the emergence of ledgers like Ethereum with
APIs of higher functionality, gave rise to various
``Decentralized Finance'' (DeFi)~\cite{DBLP:journals/corr/abs-2101-08778} applications.
Such systems have drawn the attention of industry, governments, 
regulators, and banks worldwide. 
Nonetheless, there is no agreement as to whether blockchain systems are 
decentralized, or even what ``decentralization'' entails, despite it being a 
topic of interest for centuries and across different 
disciplines~\cite{bastiat1846popular,undp1999decentralization,johnson1999diversity}.
Proponents often tout 
the existence of diverse communities, wide geographical distribution, or 
a theoretical ability of open participation as evidence of 
decentralization~\cite{antonopoulos}.
Antagonists point to 
power concentration around a few entities when it comes to system
maintenance, protocol upgrades, or wealth ownership~\cite{roubini2018lie}. 
Interestingly, both sides might be correct at the same time --- to some extent.
Blockchains may exhibit high levels of decentralization \wrt some
aspects, but not others. Thus, the pertinent question is more nuanced than 
the simple binary one ``is the system  decentralized or not?'' --- we are 
interested to know to what degree and in which aspects the system is 
(de)centralized.

Another common fallacy is perceiving decentralization as a goal, instead of a 
means to an end, and equating it with security, stability, or even efficiency.
In reality, decentralization guarantees none of these properties. It can be
synergistic to them, but in practice centralized systems can be more secure and 
fail-safe than decentralized ones and vice versa, depending on the relevant 
threat model. 
Still, it can be argued that decentralization's major advantage from a security perspective is related to 
the system's \emph{resilience to single points of failure}. 

With this as a starting point, our work sets on exploring decentralization across different 
layers, or strata, of blockchain systems\ifsubmission. \else
, systematizing a methodology for evaluating how existing systems fare \wrt 
each layer, 
and identifying relevant directions for future research. 
\fi
In particular, we select layers that influence a distributed system's security 
properties, \eg privacy or fault tolerance. 
Thus, centralization in one of our layers points to the existence of 
a single point of failure for the system as a whole \wrt one of those 
properties.

Our systematization effort aims
to inform users, practitioners, and researchers, and to support  
policymaking and law enforcement processes. 
Decentralization --- or the lack of it --- plays a major role
in policy discussions and the debate over the
regulation of blockchain systems.
\ifsubmission
For example, to determine if a digital asset constitutes a security, and
particularly an investment contract, the US Securities and Exchange Commission
(SEC) focuses on whether asset owners expect to
profit via the efforts of ``active participants'' (APs), \eg promoters or
\ifsubmission
sponsors~\cite{securities2019framework}
(see also Appendix~\ref{sec:regulation}).
\else
sponsors~\cite{securities2019framework}.
\fi
If a system
is deemed decentralized across all layers, in effect there is no AP
that the system's stakeholders rely on for profiting, so the underlying token
would not be classified as a security under this criterion.
\else

\ifsubmission
\section{Decentralization and Policymaking}\label{sec:regulation}
\fi
In the US, an early policymaking decision 
was the SEC's ruling of
``DAO'' tokens as securities~\cite{securities2017sec}. It argued
that the DAO was predominantly controlled by its creators,
who handpicked parties with decisive
operational capabilities. In $2018$, 
it was posited that characterization as a security
depended on whether the token's network is sufficiently
decentralized~\cite{hinman2018digital}. 
This was highlighted in SEC's guidance~\cite{securities2019framework}, where the
existence of ``active participants'' (APs) that undertake
essential responsibilities was named as a deciding factor for the 
classification of a digital asset as a security --- particularly an 
``investment contract.''\ifshort~\else\footnote{In $2019$, a core argument for characterizing 
XRP as a security was that token ownership, distribution, development, and 
management were controlled by a single entity~\cite{ripple2020sec}.} \fi
Our methodology, which measures decentralization as the distance from single 
points of failure in different strata, could thus aid law enforcement in making such decisions. In essence, an AP, as 
described by the SEC, is also  a single point of failure from a (cyber-) security perspective so, if a 
system is centralized under our methodology, the sale of its digital asset 
is possibly constituting an investment contract, as opposed to a system
proclaimed to be decentralized. Our minimum decentralization test
(Definition~\ref{def:mdt}) formalizes this perspective and could serve as a litmus test.

In the UK, the Financial Stability Board (FSB) identified the possibility of
power concentration around a small set of parties, when it comes to ownership
and operation of key infrastructure, as a major concern in the risks of the
application of decentralized technologies~\cite{board2019decentralised}. Nonetheless,
it also warned that decentralization
in conjunction with inadequate governance ``makes it difficult
to resolve technological limitations or errors and may lead to
uncertainty''~\cite{board2018crypto}. 
 
Decentralization is also a matter
of interest in the European Union. EU member states have suggested that
legislation should take into account the decentralized nature of the technology
on which various businesses operate~\cite{esma2019}. In late $2022$, the
Markets in Crypto-assets regulation (MiCA) was approved by the EU council and
the Parliament Committee on Economic and Monetary
Affairs~\cite{mica}.\footnote{A final vote in a full parliament session
is expected by the end of $2022$.} This
regulation makes a specific mention of decentralization as \emph{the}
distinguishing factor on whether a system falls within its scope.
Interestingly, systems might fall under this regulation even if some parts of them
are decentralized, but they are not fully decentralized, \ie across all
relevant strata.\footnote{``This Regulation applies to natural, legal persons
and other undertakings and the activities and services performed, provided or
controlled, directly or indirectly, by them, including when part of such
activity or services is performed in a decentralized way. Where crypto-asset
services as defined in this Regulation are provided in a fully decentralised
manner without any intermediary they do not fall within the scope of this
Regulation.''~\cite{mica}} 

\fi

We note that many blockchain systems can be argued to have a \emph{potential} for
decentralization, due to their permissionless nature. Specifically, by allowing
any party to join, they may find themselves in a 
decentralized state. Nonetheless, our work focuses on characterizing the
decentralization of systems as manifested in specific points in time based on the engagement they attract, thus exploring to what
degree these systems realize their decentralization potential in the real world, irrespective of whether they can be decentralized in theory.

\smallskip \noindent \textbf{\emph{Related Work.}}
Various research works have addressed the decentralization --- or 
lack thereof --- of blockchain systems, from some particular perspective.
The research of Zhou~\cite{zhou2019irregular} and Cho~\cite{cho2018asic} 
highlights the risk of centralization that arises in the context of 
\textit{hardware}, when specialized equipment is used by system maintainers to 
create blocks. This tendency is also acknowledged in the work of Ekblaw 
\etal~\cite{ekblaw2016bitcoin}. 
Choi \etal~\cite{choi2022attack} and Reibel \etal~\cite{FC:ReiYouMei19} reveal 
high levels of similarity in the codebases of different blockchain projects, 
alluding to centralization around the \textit{software} used in distributed 
ledgers. An empirical study by Azouvi \etal~\cite{FCW:AzoMalMei18} also looks 
at software centralization within a single project, \ie when 
few individuals undertake the majority of the development process.
Neudecker \etal~\cite{neudecker2018network} identify several ways in 
which the underlying \textit{network} of a distributed ledger can impact its 
overall degree of decentralization, while Apostolaki 
\etal~\cite{SP:ApoZohVan17} examine 
centralization on the level of Autonomous Systems (ASes) as an enabler of 
routing attacks on blockchains. 
A plethora of studies, such as those by Gencer \etal~\cite{FC:GBEvS18}, Gervais 
\etal~\cite{DBLP:journals/ieeesp/GervaisKCC14},
Valdivia \etal~\cite{valdivia2019decentralization} or Lin 
\etal~\cite{lin2021measuring}, have focused on the decentralization of the 
\textit{consensus} layer, by measuring the ``mining power'' ratio of a system's 
block producers. 
Another blockchain dimension whose decentralization has been thoroughly studied 
is the one pertaining to the economics of cryptocurrencies --- often termed 
\textit{tokenomics}. Sai \etal~\cite{DBLP:journals/fbloc/SaiBG21}, Cheng 
\etal~\cite{cheng2021decentralization} and Ron and Shamir~\cite{FC:RonSha13} 
analyze the distribution of transactions and tokens across parties, while Moore 
and Christin~\cite{moore2013beware} 
touch on the subject of secondary markets and the risk carried by their 
potential centralization.
Chatzigiannis \etal~\cite{chatzigiannis2022sok} point out that most blockchain 
light \textit{client} schemes are vulnerable to centralization because of their
reliance on centralized servers or full nodes, a concern also shared by Moxie
Marlinspike~\cite{marlinspike2022my}.
Gervais \etal~\cite{DBLP:journals/ieeesp/GervaisKCC14} present examples of 
centralization from the space of blockchain \textit{governance}, and 
particularly conflict resolution, while Azouvi \etal~\cite{FCW:AzoMalMei18} 
complement this work with a more systematic exploration of the contributors 
behind improvement proposals and discussions.
Various works, such as those of Mariem \etal~\cite{mariem2020all} and Sun 
\etal~\cite{sun2022spatial}, turn their attention to the \textit{geographic} 
dispersion of participants and infrastructure within a 
blockchain ecosystem.

Despite the breadth and depth of the research around blockchain 
(de)centrali\-zation and its manifestations, there have been few efforts
so far to generalize or systematize this knowledge. 
Sai \etal~\cite{DBLP:journals/ipm/SaiBFG21} offer a blockchain 
centralization taxonomy, based on an algorithmic literature review and expert 
interviews. They treat ledgers as multi-layer systems, capturing 13 aspects of 
centralization over 6 architectural layers: Application, Operational, 
Incentive, Consensus, Network, and Governance.
However, their work neglects some components, such as software centralization 
(as identified in~\cite{choi2022attack,FC:ReiYouMei19,FCW:AzoMalMei18}), 
or geographic decentralization pertaining to layers other than the network (for 
example, the decentralization of consensus participants, as studied by Sun 
\etal~\cite{sun2022spatial}). 
More recently, Zhang \etal~\cite{zhang2022sok} propose a taxonomy around five 
facets of decentralization: Consensus, Network, Wealth, Governance, and 
Transactions.
They focus primarily on transaction centralization (\wrt the distribution 
of transactions to users), which is mainly a measure of adoption and usage, 
rather than a dimension with security implications.
Their systematization also does not account for several factors 
identified in previous research, including 
hardware~\cite{zhou2019irregular,cho2018asic}, 
software~\cite{choi2022attack,FC:ReiYouMei19,FCW:AzoMalMei18}, or 
geographic~\cite{mariem2020all,sun2022spatial} decentralization.
%
Last, there exist some studies that approach the topic of blockchain 
decentralization from different perspectives, \eg economic or 
social~\cite{bodo2019logics,bodo2021decentralisation}. 
Notably, while all these works offer ample information on blockchain 
decentralization, none of them propose a consistent methodology for determining 
the decentralization level across all relevant layers.

%

\section{Methodology}\label{sec:methodology}

\begin{figure}
    \center
    \includegraphics[width=0.7\columnwidth]{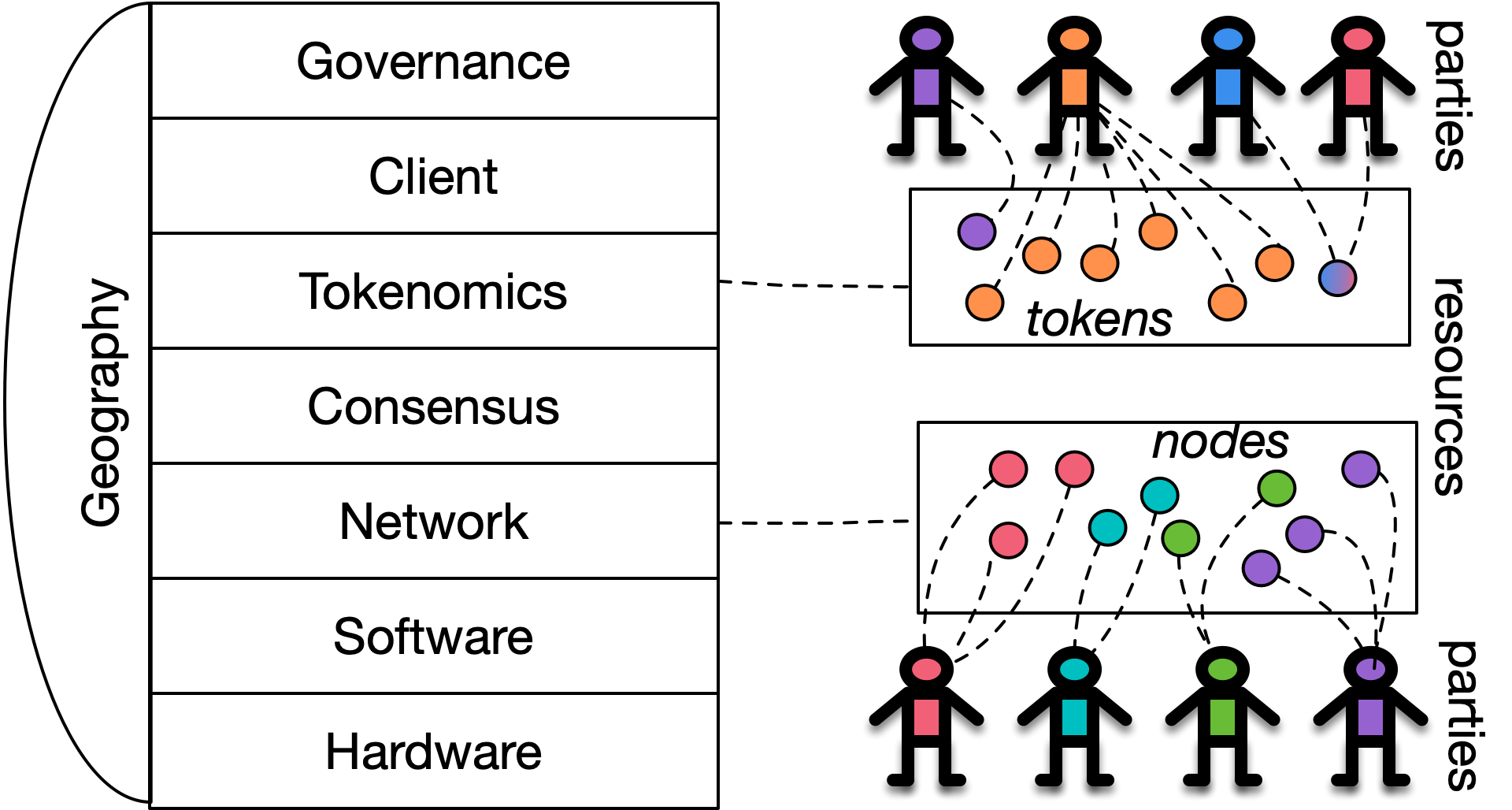}
    \caption{
    	Illustration of our methodology: the layers of a blockchain system,
        identification of some type of resources in two of the layers (tokens,
        peers), their assignment to relevant parties exhibiting a higher
        (network) and lower (tokenomics) degree of decentralization and an
        example of equal joint ownership of one token.
    }
    \label{fig:layers}
\end{figure}

%
Decentralization in the context of blockchains is often reduced to particular 
aspects of the system \eg consensus participation. 
Nonetheless, distributed ledgers comprise multiple essential,
interacting components.
Drawing from all sources of prior research, our work discerns the layers 
that form a ledger in a ``bottom-up''
manner.\footnote{Our stratification is inspired by the OSI conceptual network
model, cf. \cite{DBLP:books/lib/TanenbaumW11}.} Starting from the physical layer, \ie \emph{Hardware}, we systematize 
blockchain decentralization in multiple strata, all the way up to 
\emph{Governance}. We also include \emph{Geography} as a dimension that touches 
upon all other layers
(Figure~\ref{fig:layers}). We note that this layering 
is applied only on the ledger's stack; exploring decentralization in
exogenous infrastructure (\eg physical links, Internet routing, operating
systems \etc) is an interesting question, but outside the scope of this paper.

A first step to understand the importance of decentralization for such systems 
is to identify the properties of interest that 
distributed ledgers should satisfy and which can be affected by the ledger's 
degree of decentralization. 
The two core security properties that each ledger should guarantee are 
\emph{safety} and 
\emph{liveness}. Safety ensures that all honest users hold the same,
``settled'' view of the ledger. Liveness reflects the ability to update the
ledger's settled view regularly, as new transactions are submitted.
We note that safety and liveness typically incorporate and express other useful
properties for real-world applications, such as transaction finality or 
censorship resistance. 
A third property, \emph{privacy}, guarantees that
users' actions enjoy a certain degree of dissociation from their real-world identities
and individually are unlinkable. 
Finally, specifically in the context of blockchain systems,
\emph{price stability} captures the property that the ledger's core
digital asset's supply and market price are predictable (to a reasonable
extent). In particular, price stability is violated if the asset's market price
demonstrates high volatility in the short term (\eg 
monthly).\footnote{The cryptocurrency market is notoriously volatile, so one could apply different thresholds for ``reasonable'' levels of volatility. An interesting line of future research would be to identify non-cryptocurrency assets, which could serve as a base of comparison for which levels of volatility are acceptable in this case.}
For ease of reading, we will refer to this property only as ``stability'' for the rest of the paper.
 
We view these properties through the lens of (cyber-)security, 
\ie in the context of an adversary who wishes to subvert them. 
This is strictly stronger compared to settings where failures are assumed to be
benign (\eg crash faults due to power outages).
A single point of failure exists when a single party, if
controlled by the adversary, can violate one or more of the ledger's properties. 

We lay out the following methodology: 
for each system layer we identify
\begin{inparaenum}[(a)]
    \item one or more \emph{resources}, that can be thought of as the basic
        ``unit'' of the layer pertaining to the ledger's security
        properties;
    \item the \emph{relevant parties} that control, either directly or
        indirectly, said resources;
    \item the ledger's \emph{properties} that are at risk, if the
        resources' distribution across the relevant parties becomes
        centralized.
\end{inparaenum}
For example, considering Bitcoin's consensus layer, the resource is hashing power and 
the relevant parties are the miners; the properties at risk are
safety, liveness and, to a somewhat lesser degree, stability and privacy.
\ifsubmission\else

We employ this framework for each layer, along with relevant 
evaluations from the literature regarding deployed systems.\footnote{The lack 
of public data \wrt some dimensions makes it impossible to evaluate the same 
set of systems throughout our paper, hence, we present findings depending on 
the available information. As the most studied systems, Bitcoin and Ethereum 
are included in most evaluations, with others (\eg Solana, Cardano) evaluated 
when possible.} The main benefit of our systematization effort is thus laying 
out the foundations for assessing any given system's decentralization level in 
a quantitative manner. 
\fi
Table~\ref{tab:summary} provides a summary of our systematization, with 
resources and relevant parties presented for each identified layer and 
sub-category.

Notably, a resource might be modular, with some parts 
considered more important than others. For instance, software products are 
typically not monolithic, with \eg documentation being less crucial than a 
library or a configuration file. Therefore, the parties that maintain the 
former have (arguably) less influence over the resource (\ie the software 
product) than the coders of the latter. To resolve this concern, one could 
compute an aggregate level of decentralization, after \emph{weighing} each 
component based on its significance. Such aggregation 
methodology 
could also be applied to compute the decentralization level of the whole 
system, assuming weights for each layer (cf. Section~\ref{sec:conclusion}).

Another issue is that one relevant party may encompass multiple 
real-world identities. For instance, consider two software products, one 
maintained by a single organization with many members, the other maintained by 
a handful of independent developers. Although the first may be more 
decentralized in terms of people, from a legal perspective the
second may be deemed more decentralized, as the first is authored by a single legal entity.

\begin{table*}[!ht]
    \centering \def\arraystretch{1.3}
    \begin{tabular}{cccc}

          \textbf{Layer}
        & \textbf{Subcategory}
        & \textbf{Resources}
        & \textbf{Relevant Parties}
        \\
        \specialrule{1.6pt}{1pt}{1pt}

        \multirow{2}{*}{\begin{tabular}[c]{@{}c@{}} 
        \textbf{Hardware} \\ 
        \cite{zhou2019irregular,cho2018asic} \end{tabular}}
        & Physical hardware 
        & Participating power
        & \begin{tabular}[c]{@{}c@{}} Hardware manufacturers 
        \end{tabular}
        \\
        \cdashline{2-4}

        & Virtual hardware
        & Participating power
        & Cloud providers
        \\
        \hline

        \ifshort
        \multirow{2}{*}{\begin{tabular}[c]{@{}c@{}} 
        		\textbf{Software} \\ 
        		\cite{choi2022attack,FC:ReiYouMei19,FCW:AzoMalMei18}
        		 \end{tabular}}
        \else
        \fi
        & Protocol participation
        & \begin{tabular}[c]{@{}c@{}} 1) Participating power \\ 2) Full nodes 
        \end{tabular}
        & \begin{tabular}[c]{@{}c@{}} 1) Full node software \\ implementations 
        \\2) Developers of \\ full node software 
        \end{tabular}
        \\
        \cdashline{2-4}

        \ifshort
        \else
        \textbf{Software}
        \fi
        & Asset management
        & Tokens
        & \begin{tabular}[c]{@{}c@{}} 1) Wallet \\ implementations 
        \\2) Developers of \\ wallet software
        \end{tabular}
        \ifshort
        \\
        \hline
        \else
        \\
        \cdashline{2-4}
        \fi

        \ifshort
        \else
        & Testing
        & Testnets
        & Development teams
        \\
        \hline
        \fi

        \multirow{2}{*}{\begin{tabular}[c]{@{}c@{}} 
        		\textbf{Network} \\ 
        		\cite{neudecker2018network,SP:ApoZohVan17,FC:GBEvS18,DBLP:conf/iscc/0003SWTZY19}
        \end{tabular}}
        & Topology
        & Component bridges
        & Owners of bridges
        \\
        \cdashline{2-4}
        & Peer discovery
        & Bootstrapping nodes
        & Node operators
        \\
        \hline

        \begin{tabular}[c]{@{}c@{}} 
        	\textbf{Consensus} \\ 
        	\cite{FC:GBEvS18,DBLP:journals/ieeesp/GervaisKCC14,valdivia2019decentralization,lin2021measuring}
        \end{tabular}
        & 
        & \begin{tabular}[c]{@{}c@{}} 1) Owned participating power \\ 2) 
        Delegated participating power \end{tabular}
        & \begin{tabular}[c]{@{}c@{}} 1) Owners of \\ participation nodes \\2) 
        Pool leaders and delegates
        \end{tabular}
        \\
        \hline

        & Initial distribution
        & Bootstrapping tokens
        & Token holders
        \\
        \cdashline{2-4}

        \begin{tabular}[c]{@{}c@{}} \textbf{Tokenomics} \\ 
        \cite{DBLP:journals/fbloc/SaiBG21,cheng2021decentralization,FC:RonSha13,moore2013beware}
        \end{tabular}
        & Token ownership
        & Tokens
        & \begin{tabular}[c]{@{}c@{}} 1) Key managers \\ 2) Legal stakeholders 
        \end{tabular}
        \\
        \cdashline{2-4}

        & Secondary markets
        & Market liquidity
        & \begin{tabular}[c]{@{}c@{}} 1) Exchanges \\ 2) Trading pairs \end{tabular}
        \\
        \hline

        \begin{tabular}[c]{@{}c@{}} \textbf{Client 
        API}\\ \cite{chatzigiannis2022sok,marlinspike2022my} 
        \end{tabular}
        &
        & Tokens
        & Full node operators
        \\
        \hline

        \multirow{2}{*}{\begin{tabular}[c]{@{}c@{}} 
        \textbf{Governance} \\ 
        		\cite{DBLP:journals/ieeesp/GervaisKCC14,FCW:AzoMalMei18,DBLP:journals/ipm/SaiBFG21}
        \end{tabular}}
        & Conflict resolution
        & Decision-making power
        & \begin{tabular}[c]{@{}c@{}} All system entities \end{tabular}
        \\
        \cdashline{2-4}

        & R\&D funding
        & Capital
        & Active developers
        \\
        \midrule[1pt]

        \multirow{2}{*}{\begin{tabular}[c]{@{}c@{}} 
        \textbf{Geography} \\ 
        		\cite{mariem2020all,sun2022spatial,DBLP:journals/ipm/SaiBFG21}
        		 \end{tabular}}
        & Physical safety
        & All resources above
        & Regions
        \\
        \cdashline{2-4}

        & Legal compliance
        & All resources above
        & Jurisdictions
        \\
        \hline
        \\
    \end{tabular}
    \caption{
        Overview of blockchain decentralization layers, 
        including for each layer the literature that 
        motivated it and the way it fits into our framework.
    }
    \label{tab:summary}
\end{table*}

By projecting the relevant parties of each category to \emph{legal persons},
we articulate a test that can be useful in assessing systems \wrt their
decentralization in a legal sense (Definition~\ref{def:mdt}). Here, a
legal person can be 
\ifshort
an organization, \eg a company or non-profit foundation, or an individual.
\else
a company (\eg a foundation that performs the
development of the system), but also an individual (\eg a so-called ``whale'',
an individual who possesses a significant number of tokens). We note that,
although in this work the MDT is considered \wrt legal parties, it could be
adapted to treat exogenous centralization aspects as well, \eg relating to natural disasters.
\fi

\begin{definition}\label{def:mdt}
    A blockchain system fails the \emph{Minimum Decentralization Test} (MDT) if
    and only if there exists a layer (cf. Table~\ref{tab:summary}) for which
    there is a single \emph{legal person} that controls a sufficient number of
    \emph{relevant parties} so that it 	is able to violate a property of
    interest.
\end{definition}

In the following sections, we provide detailed explanations 
as to why each identified layer is important and how 
it fits into our framework 
(Sections~\ref{sec:hardware}-\ref{sec:geography}). Then, we 
apply our methodology on case studies 
(Section~\ref{sec:case-studies}), and finally, we suggest 
various directions for future research that our work points 
to (Section~\ref{sec:conclusion}).

\section{Hardware}\label{sec:hardware}

The role of hardware in the potential decentralization of blockchain systems 
has been reported in various research 
works~\cite{zhou2019irregular,cho2018asic}.
To be specific, by ``hardware'', we refer to the machines that host and/or support
the consensus software, which can be anything 
from personal computers to purpose-built devices.
In many cases, the hardware is also provided as a service from cloud providers 
to consensus participants.
To account for all possibilities, we  segment this layer into two categories, 
namely ``physical'' and ``virtual'' hardware.

\smallskip \noindent 
\textbf{\emph{Physical hardware.}}\label{subsec:physical-hardware}
This category covers all hardware that is used directly by consensus 
participants.
Bitcoin mining, and that of other Proof-of-Work (PoW) ledgers, started from 
regular CPUs, but quickly migrated to GPUs and, eventually, to dedicated 
devices (ASICs), which produce more hashes at a lower 
cost~\cite{taylor2017evolution}.
In the case of Proof-of-Stake (PoS) blockchains, participation is typically 
performed through generic hardware, although there are systems with 
particular requirements, which may restrict the compatible hardware options (an 
example such blockchain is The Internet Computer by 
Dfinity~\cite{dfinity_requirements}).
Some alternative schemes, such as certain forms of Proof-of-Useful-Work (PoUW) or 
Proof-of-Elapsed-Time (PoET) also make use of Trusted Execution Environments 
(TEEs) to guarantee higher security or efficiency 
levels~\cite{bao2020blockchain}.

When analyzing decentralization on this layer, the \emph{resource} of interest  
is participating power (\eg hashes per second or stake) and the \emph{relevant 
parties} are the manufacturers of hardware products that are used for 
participating.
That is, we consider a system to be decentralized in the hardware layer if the 
participating power is distributed across various pieces of equipment,  
manufactured by different entities.

As demonstrated by the evolution of Bitcoin mining, one pathway to  
hardware centralization is the development and adoption of specialized 
machines that outperform generic equipment and provide an advantage to their operators. 
Nowadays, an overwhelming majority of block production in PoW blockchains comes 
from specialized hardware, despite the fact that PoW does not require 
specialized hardware in theory~\cite{taylor2017evolution}.
Notably, this trend has motivated significant research in ``ASIC-resistance'' 
and the development of PoW algorithms that attempt to facilitate better 
hardware diversity~\cite{argon2,ethash,TCC:RenDev17}.
In other cases, strict or ``non-typical'' protocol requirements (\eg the 
requirement for trusted hardware) reduce the pool of possible manufacturers 
that can support those systems, potentially leading to increased 
centralization. 

Concentration around few hardware manufacturers creates various hazards.
Same-vendor products are more susceptible to collective faults, \eg due to 
defective parts or hardware bugs. 
Such faults could result in sudden drops in the network's
power, lowering the threshold for gaining a computational majority
(\emph{safety} and \emph{liveness} hazard) and slowing down block production, 
at least
until the PoW parameters are recalculated (\emph{liveness} hazard).
Manufacturers could also introduce backdoors, threatening the
ledger's security and \emph{stability}, albeit such hazards can possibly be
mitigated via cryptographic techniques~\cite{PROVSEC:AKMTV18}.

\ignore{
In theory, PoW allows anyone to participate, regardless how small an amount of 
power they control. 
In practice, some machines are more efficient in solving the PoW challenge 
than others, hence, miners who use such machines have a competitive advantage. 
Such shift to dedicated devices introduces various centralization tendencies.  
First, producing such devices requires a high-level of expertise and is an
investment in the blockchain system itself. In addition, microchip
manufacturing observes economies of scale, since the barrier to enter the
system in terms of designing and packaging the circuits is high. Consequently,
the market tends to concentrate around few companies.
Moreover, dedicated devices are particularly expensive,\footnote{As of $2022$,
popular Bitcoin ASICs cost (tens of) thousands of USD (\eg 
\href{https://shop.bitmain.com/}{Bitmain}).} 
but significantly more profitable than generic hardware. 
Industrial farms host hundreds of ASICs, 
resulting in a profitability gap between ASIC and generic
hardware users~\cite{karakostas2019cryptocurrency}. 
}

\ifshort\else
\ifshort
\else
\subsubsection*{Evaluation}
\fi
Centralization around specialized hardware has been
documented~\cite{ekblaw2016bitcoin}, although no academic research could be
found on mining hardware usage in real-world systems. Interestingly, it is
unclear how to even measure the usage of hardware equipment in PoW mining via public data, as
well as how to develop PoW algorithms that promote diversity, thus future
research could aim at answering these questions.
Nonetheless, there exist some reports,
though they often present conflicting assessments. In Bitcoin,
between $2017-2019$ a single mining hardware provider accounted for either
$65-75$\%~\cite{bendiksen2019bitcoin} or $46-58$\%~\cite{tokeninsight-2} of the
network's hashrate, with $98$\% of the market controlled by $4$ firms.


\fi

\smallskip \noindent 
\textbf{\emph{Virtual hardware.}}\label{subsec:virtual-hardware}
The emergence of mining data centers allowed for hashing power to be offered as 
a ``cloud'' service, effectively enabling miners to participate in block 
production without possessing their own 
hardware~\cite{magaki2016asic,tosh2017security}.
The advent of PoS protocols reinforced this trend towards 
``virtual'' hardware, by decoupling Sybil resilience from physical 
requirements. 
In theory, this enables PoS nodes to run on generic hardware, \eg even home 
equipment, but in practice, convenience often drives PoS users to employ cloud 
services, which offer uptime and connectivity guarantees that a DIY 
configuration cannot. 
This is exacerbated when PoS systems apply penalties to absent users and 
uptime guarantees become of utmost importance to guarantee profitability. 
Therefore, the \emph{resource} of interest in this category is again 
participating power (either in the form of hashing power or stake) and the 
\emph{relevant parties} are cloud providers.

When nodes that control significant participating power are hosted by the same provider,
significant hazards arise for all properties. First, the provider may have access to private keys and hence is able to create
conflicting blocks (\emph{safety} hazard) or deanonymize users (compromising 
\emph{privacy}).
Second, the provider controls the node's network access, so it could prohibit
communication (\emph{liveness} hazard). Finally, it could also tamper with the
system's \emph{stability}, \eg increasing price volatility via targeted
interference or even stealing user rewards.

\ifshort\else
\ifshort
\else
\subsubsection*{Evaluation}
\fi
A comprehensive evaluation of centralization in terms of hardware hosting, and 
how to incentivize hosting diversity is a promising thread of future research.
Here, we consider two examples of highly-valued PoS
systems, Solana and Avalanche.\footnote{\href{https://solana.com}{Solana} and
\href{https://www.avax.network/}{Avalanche} are \#9 and \#14 respectively \wrt
market capitalization. [\href{https://coinmarketcap.com}{CoinMarketCap}; August
2022]} 
Solana's validators predominantly operate cloud-based nodes; of the $1873$
nodes, more than half are hosted in two services, with more than $50$\% and
more than $66$\% of participating stake hosted by $3$ and $5$ providers
respectively.\footnote{\href{https://www.validators.app/data-centers?locale=en\&network=mainnet\&sort\_by=asn}{validators.app}.
[August $2022$]}
Avalanche observes similar concentration issues; $731$ out of $1254$ validators,
who control $71.84$\% of all stake, are hosted by a single
company.\footnote{Data obtained from
\href{https://avascan.info/stats/staking}{avascan.info}. [August $2022$]} 


\fi

\section{Software}\label{sec:software}
Software development is another dimension of distributed ledgers 
that has been associated with potential 
centralization~\cite{choi2022attack,FC:ReiYouMei19,FCW:AzoMalMei18}.
Diverse software development and usage is a core element of stability and
safety of distributed ledgers, as it increases resilience to
catastrophic bugs in a product's code. A vulnerability in
one implementation may jeopardize a part of the system, but if the system is 
sufficiently decentralized, such vulnerabilities would not escalate to systemic threats.
Following, we discuss the development of core blockchain software components,
namely transaction validation and PoW mining (via full nodes) and management of
keys and digital assets (via 
wallets).\footnote{Appendix~\ref{subsec:software-testing} also explores 
software testing.}

\smallskip \noindent \textbf{\emph{Protocol 
Participation.}}\label{subsec:full-nodes}
The principal type of software in blockchain systems is the \emph{full node}.
Full nodes implement the ledger protocol by: 
\begin{inparaenum}[i)]
    \item keeping a local chain;
    \item validating new transactions;
    \item extending the local chain with new blocks;
    \item participating in the consensus mechanism to incorporate new blocks.
\end{inparaenum}
To analyze decentralization, we identify two \emph{resources} of interest:
\begin{inparaenum}[1)]
    \item (number of) full nodes;
    \item participating power (\eg computational or stake) that is hosted on
        full nodes.
\end{inparaenum}
The \emph{relevant parties} are:
\begin{inparaenum}[1)] 
	\item full node software implementations;
	\item full node software developers. 
\end{inparaenum}

Relying on a handful of full node implementations introduces
\emph{safety}, \emph{liveness}, and \emph{stability} hazards. A
bug that fails to validate correct transactions would hurt the system's
liveness, whereas accepting incorrect transactions could hurt the system's
safety and stability, \eg via a network split or token forgery. Such bugs have been observed 
in Bitcoin Core and could have resulted in
Denial-of-Service (DoS) attacks~\cite{bitcoin-bug-2} and
token forgery~\cite{bitcoin-bug-1}. Implementation bugs could
also threaten \emph{privacy}, if nodes reveal information
about message origin (\eg IP addresses). 
Similar threats arise when code gets reused across different projects. Often, a 
new blockchain is only a ``derivative'', that is a project that started by 
forking an existing codebase, including copyrighted 
information~\cite{FC:ReiYouMei19}. 
Such projects often remain unpatched, so bugs in the initial implementation
tend to spill over~\cite{choi2022attack}. Vulnerabilities may also arise when
adapting an existing implementation in a new setting, \eg copying Bitcoin's
code and replacing PoW with PoS~\cite{FC:KKLM19}.
Thus, widespread usage of
multiple node implementations, developed by different teams, is a
hallmark of a secure and reliant ecosystem.

\ifshort\else
\ifshort
\else
\subsubsection*{Evaluation}
\fi
The literature is lacking formal analyses on the usage of full node
ledger software, so a rigorous evaluation of the dynamics in software development
and usage could highlight various centralization tendencies.
%
Various community and commercial projects do keep track of
statistics though. In most systems, a single client software is
predominantly used by the participating nodes in the network. In Bitcoin
$99$\% use Bitcoin Core (aka Satoshi), in Ethereum $78$\% use geth, in
Litecoin $95$\% use LitecoinCore, while systems like Zcash are completely
centralized with all nodes using one software (MagicBean); a notable exception
is Bitcoin Cash, where usage is split between BCH Unlimited ($33$\%), Bitcoin
Cash Node ($51$\%), and Bitcoin ABC ($12$\%).\footnote{Sources:
\href{https://blockchair.com}{blockchair},
\href{https://ethernodes.org/}{ethernodes} [August 2022]} 


Some projects are actively managed by a wide network of
developers, \eg more than $200$ contribute to Ethereum~\cite{devtrends}, while others are particularly
centralized. As of $2018$, $7$\% of all Bitcoin Core files were written by the
same person, while $30$\% of all files had a single author. In Ethereum,
these figures rise to $20$\% and $55$\% respectively~\cite{FCW:AzoMalMei18}.
Comments observe similar centralization patterns, with $8$ ($0.3$\%) and $18$
($0.6$\%) people contributing half of all comments in Bitcoin and Ethereum
respectively~\cite{FCW:AzoMalMei18}.

\fi

\smallskip \noindent \textbf{\emph{Asset 
Management.}}\label{subsec:wallets}
Distributed ledgers are mostly used for bookkeeping of digital asset
transactions, so securely managing and transferring assets is a core
necessity. Digital assets are typically managed by private keys and
represented via addresses. The software responsible for managing keys and
addresses is the wallet~\cite{karantias2020sok} and its principal
functionalities are:
\begin{inparaenum}[i)]
    \item store the user's keys;
    \item prove ownership of the assets (managed by the keys);
    \item issue transactions that transfer assets to other accounts;
    \item retrieve the user's (keys') balance and history information.
\end{inparaenum}
Therefore, the \emph{resource} of interest is the set of all assets managed via
the ledger and the \emph{relevant parties} are, accordingly:
\begin{inparaenum}[i)]
	\item wallet implementations;
	\item developers of wallet software.
\end{inparaenum}

The wallet is a major point of security consideration, so multiple properties rely
on it. A bug which \eg corrupts the user's keys could not only prohibit a user
from transacting with their assets (\emph{liveness} hazard), but forever lose access
to them (\emph{stability} hazard). This was demonstrated in $2017$, when
the ``Parity'' Ethereum wallet saw a vulnerability that allowed a user to take
ownership of multiple assets and then lock them~\cite{parity-bug}; as a result,
$300$m worth of Ethereum tokens were forever lost. In another example, some
Bitcoin wallets possibly displayed incorrect balance, effectively enabling
a double spending attack~\cite{bitcoin-wallet-bug}. Consequently, if a few
implementations are predominantly used, a vulnerability could result in
assets being unusable or stolen. Such
vulnerability could also turn into a systemic point of failure, with the whole
ledger becoming unusable.\footnote{A prime such example was ``The DAO'', which
in $2016$ attracted nearly $14$\% of all Ethereum tokens and, when hacked,
instigated a change in Ethereum's consensus layer and a hard fork which split
the network~\cite{secDAO}.}

\ifshort\else
\ifshort
\else
\subsubsection*{Evaluation}
\fi
As keys and addresses are wallet agnostic, it is
impossible to identify if two addresses are generated by the same wallet
implementation, unless it purposely reveals such information.  Consequently, it
is unclear how to evaluate the wallet market's diversity and how widespread
wallet usage is from public data. To our knowledge, no rigorous investigation has been conducted
on this topic, either analyzing historical data patterns or conducting
usability studies.
%

\ifsubmission\else
\ifsubmission
\section{Software Testing}\label{subsec:software-testing}
\else
\subsection{Testing}\label{subsec:software-testing}
\fi

Testing is a core part of software development. In blockchains,
a major means of testing new applications or ledger features is testnets. A
testnet is a separate chain, identical to the main chain in terms of offered
functionalities. 
To transact, a user acquires testnet tokens for free, so the
native testnet tokens have no real-world value. 
Testnets offer multiple functionalities. Users can test
features without risking losing funds.
Developers test new features and applications in a
scale that closely resembles the main chain.
Adversaries evaluate the efficacy of
attacks~\cite{ropsten-attack} or exploit the zero-cost nature of testnet
transactions~\cite{FC:FraAbeDaz20}. Therefore, testnets indirectly
safeguard all ledger properties.
Fewer testnets increase centralization around
specific full node software products, while
testnets maintained by diverse teams may collect richer
data. Hence, the \emph{resource} is testnets and the
\emph{relevant parties} are their operators.

\ifshort
\else
\subsubsection*{Evaluation}
\fi
Bitcoin offers a single primary testnet; the same holds for alternative cryptocurrencies like Zcash
and Monero.\footnote{Sources: \href{https://en.bitcoin.it/wiki/Testnet}{Bitcoin
Wiki},
\href{https://zcash.readthedocs.io/en/latest/rtd_pages/testnet\_guide.html}{Zcash
Docs}, \href{https://monerodocs.org/infrastructure/networks/}{Monero Docs}
[August 2022]}
In Ethereum, although seemingly multiple testnets exist, most are 
deprecated due to the system's transition to PoS, and only one 
of the recommended networks is expected to be maintained in the 
long term.\footnote{The deprecation of Ropsten, Rinkeby, Kiln and Kovan 
was announced in $2022$ 
[\href{https://blog.ethereum.org/2022/06/21/testnet-deprecation/}{Ethereum 	
blog}]. Goerli will be maintained in the long run, while the future 
of Sepolia is undecided. 
[\href{https://ethereum.org/en/developers/docs/networks/}{ethereum.org}; August 
2022]} PoS testnets, \eg in Cardano and Solana, are also highly 
centralized.\footnote{Cardano and Solana testnets are run by Input Output and 
the Solana Foundation resp.
[\href{https://testnets.cardano.org/}{Cardano Testnets},
\href{https://docs.solana.com/clusters}{Solana Docs}; August 2022]}

\fi

\fi

\section{Network}\label{sec:p2p}

Blockchain nodes communicate over a peer-to-peer (P2P) network, the
decentralization properties of which have been of great interest to 
researchers~\cite{neudecker2018network,SP:ApoZohVan17,FC:GBEvS18,DBLP:conf/iscc/0003SWTZY19}.
Systems often implement a message \emph{diffusion}
mechanism~\cite{EC:GarKiaLeo15} via a gossip protocol that avoids full graph
connectivity~\cite{DBLP:conf/p2p/DeckerW13}. Users typically use the Internet
to access the P2P overlay, though some efforts try to introduce an independent
infrastructure~\cite{blockstreamSattelite}. Here we explore networking
aspects which present single points of failure, in terms of topology and
bootstrapping. A notable research question that arises organically from our
analysis, and touches upon both following subsections, is creating a P2P
network that is both permissionless and Byzantine resilient, \eg as explored
in~\cite{DBLP:journals/iacr/CorettiKMR22,DBLP:conf/crypto/MattNT22}.

\noindent\emph{\textbf{Topology.}}
The first networking aspect of interest is the network's topology.  Evaluating
a real-world network's clustering properties is a well-known problem,
traditionally done by generating random graphs and comparing the expected with
the observed
values~\cite{erdos1960evolution,newman2003structure,albert2002statistical}.
In blockchain systems, every node maintains a list of peer connections.
Crucially, message provenance is not typically provided, so no party can know
the network origin of an incoming message.  Therefore, the \emph{resource} of
interest are ``bridges'' (single nodes or small cluster of nodes) between the
network graph's components and the \emph{relevant parties} are the bridges' 
owners or operators. Here, a component is a single node or a cluster of
tightly interconnected nodes.

A distributed network, in the tradition of Baran~\cite{baran1964distributed}, is
key in maintaining \emph{safety} and \emph{liveness}. Under the CAP
theorem~\cite{brewer2012cap}, any networked system can satisfy at most two of
the following properties:
\begin{inparaenum}[i)]
    \item a consistent data copy;
    \item data availability;
    \item network partition tolerance.
\end{inparaenum}
Ledger systems are no exception. Each node needs to maintain at least one
connection to an honest party to receive all messages and avoid eclipse
attacks~\cite{USENIX:HKZG15}. 

If some parties cannot communicate with the
rest of the network, either they halt and any transactions they submit are 
dropped (violating \emph{liveness}) or they produce
separate ledger versions (violating \emph{safety}).\footnote{In longest-chain
protocols, like Bitcoin, miners keep producing blocks in isolation, ending up
with different ledger versions. In BFT-style protocols, like Algorand, the
ledger cannot be updated if a large number of block producers become
unreachable and thus do not adopt new transactions.}
By preventing communication between a node and the rest of the network, an
attacker reduces the honest computational power and isolates that node, making
a $51$\% or a double spending attack (against the isolated node) easier to
deploy.
Additionally, an adversary can violate \emph{liveness} by blocking the node's
transactions from reaching the rest of the network.
Third, an adversary that controls all of a node's connections can link
transactions to the specific node, correlate the user's addresses, and
associate them with real-world information, such as an IP
address, thus compromising \emph{privacy}\footnote{Note that protecting privacy
in the network layer can be a particularly challenging problem, even in the
setting where no single adversary controls all of a node's connections.}.
This also holds on a macroscopic level, \ie a node that acts as a central
communication hub between two clusters can obtain information and even
deanonymize some participants (violating \emph{privacy}).

\noindent\emph{\textbf{Node Bootstrapping and Peer Discovery.}}
Joining a ledger's network and synchronizing with it is the so-called
``bootstrapping'' process. All real-world ledgers rely on an initial (trusted)
setup, the first (``genesis'') block.\footnote{Some proposals rely on
computational assumptions instead of a trusted setup~\cite{RSA:GarKiaPan20},
but their real-world performance and applicability is still untested.}
Obtaining the correct genesis block is done in an out-of-band, typically secure
mechanism, since it is often well-known and easy to validate from various
sources. 

Initially, the node needs to connect to some peers and receive all available
chains. Therefore, the \emph{resource} here is bootstrapping nodes and the
\emph{relevant parties} are these nodes' operators. Then, using the ledger
protocol's chain resolution mechanism, it decides which chain to adopt.  There
are two points of interest here.

First, it should be guaranteed that the node connects to at least one honest
peer, to retrieve all available information and avoid getting eclipsed.
As explained above, if nodes gets eclipsed, the system's \emph{liveness},
\emph{safety} and \emph{privacy} properties may get compromised.
However, connecting to honest peers is not straightforward, given that the node
has no knowledge about the network participants, a standard problem in P2P
networks~\cite{dickey2008bootstrapping}. Blockchain systems predominantly use
either hardcoded peer lists or DNS seeding~\cite{CCS:LoeQua19}, although both
techniques are censorship-prone. Notably,~\cite{CCS:LoeQua19} showed that more
secure alternatives, such as ZMap~\cite{USENIX:DurWusHal13}, cannot be
feasibly used in existing blockchain systems.

Second, even if the node connects to some honest peers, catching up by using
only genesis (``bootstrapping from genesis'') is not always feasible.
Especially in PoS systems, an attacker can effortlessly assemble an
arbitrarily-long, seemingly correct
chain (violating 
\emph{safety})~\cite{EPRINT:GazKiaRus18,buterin2014stake,li2017securing}.
To counter such attacks, some ledgers employ
checkpoints~\cite{C:KRDO17,EC:DGKR18,FC:DaiPasShi19}, which are often issued
centrally and are either hardcoded or received from the peers. Other solutions
do exist, \eg analyzing block density and relying on key
erasures~\cite{CCS:BGKRZ18} or using VDFs~\cite{C:BBBF18,DBLP:conf/fc/DebKT21},
but enforcing and/or
relaxing such assumptions still poses an interesting research problem.

%

\section{Consensus}\label{sec:consensus}

A key element of any distributed ledger is its consensus protocol, and a lot of 
the research behind blockchain decentralization has been dedicated to this 
layer~\cite{FC:GBEvS18,DBLP:journals/ieeesp/GervaisKCC14,valdivia2019decentralization,lin2021measuring}.
Protocols in our context are ``resource-based'', i.e., they are executed
by parties possessing units of an underlying resource (\eg hashing power or stake).
To guarantee safety and liveness, at least a
majority (or in some cases a supermajority of $\frac{2}{3}$) of the ledger's participating resources
should be honestly controlled~\cite{RSA:GarKia20}. Therefore, when a handful of actors
control enough resources to break one of the properties, a direct point of 
failure arises.

Protocols like Bitcoin~\cite{EC:GarKiaLeo15}, Algorand~\cite{EPRINT:CGMV18}, or
Ouroboros Praos~\cite{EC:GarKiaLeo15}, enable resource holders to engage in the
protocol directly with (essentially) whatever amount of resources they have.
In these protocols, block producers can, even though they do not have to, form 
coalitions called \emph{pools}. 
In PoW, a pool ``leader'' validates transactions, and organizes
them in a candidate block, while each ``member'' executes the PoW puzzle for
the leader-made block. If a member is successful, the leader collects the
block's reward and distributes it, proportionately to each member's power. In
PoS, the leader has full control over the block's creation, while the members
only pay fees to delegate their staking rights to the leader and collect 
rewards.  Pooling
behavior is also driven by temporal discounting~\cite{Reed2011}, \ie the
tendency to disfavor rare or delayed rewards. In essence, a small miner may
prefer small frequent payments, at the cost of some fee, over rare
large payments, when producing a block. 

Other systems, like Cosmos~\cite{kwon2019cosmos} and
EOS~\cite{io2018eos}, impose restrictions on which parties can
participate in consensus and require the rest to delegate their resources to a
representative or ``validator'' node. This ``barrier to entry'' means that any party without
enough stake, \ie below the system's threshold or less than its competitors, is
required to delegate their staking rights to a validator. At every ``epoch'', a
committee of (a fixed number of) parties is elected to run the protocol. The
election mechanism is voting-based, with resource delegation acting as the
voting process. 


In both types of systems, there are two \emph{resources} of interest: 
\begin{inparaenum}[i)]
    \item owned participating power, \eg computational or stake;
    \item delegated participating power, including the power to choose
        a block's content.
\end{inparaenum}
Accordingly, the \emph{relevant parties} are:
\begin{inparaenum}[i)]
	\item miners and stakeholders, who own hashing power and stake respectively;
	\item pool leaders and delegates, who control how the resources are used.
\end{inparaenum}

Typically, the security of a ledger is guaranteed if the parties that represent 
an aggregate majority of the participating power are honest (\ie they follow 
the protocol as prescribed)~\cite{EC:GarKiaLeo15}.
Therefore, the concentration of participating power around few entities poses a 
threat to the system.
This hazard is well-known and blockchain users and participants have actively
tried to avoid it since at least $2014$~\cite{coindesk-ghash}.
Those controlling a power majority can hurt \emph{liveness}, by refusing to 
publish or accept certain transactions, as well as \emph{safety} by launching a
long-range attack. Both types of attacks also indirectly hurt \emph{stability},
since the system's trustworthiness is challenged.

A second concern revolves around block proposers. A proposer is a
party that maintains a mempool and chooses which transactions are added to a
block and in what order. Initially, a single party acted as both block
proposer and builder. With the increase in hardware requirements needed to run
a full node and the formation of pools, the two roles of proposer and builder
were separated.

In PoW ledgers, the leader of the pool typically proposes the block's content,
whereas the pool members only run the PoW algorithm. Therefore, pool members
are not involved in a block's construction and often do not even validate its
contents.  Therefore, the leader may censor transactions (\emph{liveness}
hazard), steal member rewards (\emph{stability} hazard), or possibly link the
user's resources with information like IP addresses (\emph{privacy} hazard).

In addition, smart contracts enable MEV-type
attacks~\cite{DBLP:conf/sp/ZhouXECWWQWSG23}, which might hurt stability. Here,
block builders have the ability to observe transactions before publication and
choose their order in a block, which they can exploit to
extract value from honest transactions. A countermeasure that has been 
introduced is the proposer-builder separation (PBS) model, wherein a trusted
party maintains a mempool and proposes a block, whereas validators sign it
without ever observing its content (thus not being able to exploit its 
MEV)~\cite{buterin2021PBS}. Still, the current implementation of PBS in 
Ethereum has been criticized for facilitating censorship and 
centralization, hence its usefulness remains 
unclear~\cite{heimbach2023ethereum}.

Finally, a threat arises due to the lack of self-healing, \ie the inability to
recover from a temporary adversarial takeover. In PoW, even if a majority gets
corrupted, honest users can increase their own power and, eventually, overthrow
the adversary and restore the ledger's
security~\cite{DBLP:conf/fc/AvarikiotiK0W19,EPRINT:BGKRZ20}. In PoS though,
power shift takes place on the ledger, by transferring stake. If an adversary
temporarily obtains a  majority, they can prohibit transactions that shift
power away from them, thus retaining control indefinitely (for example, a large
centralized cryptocurrency exchange can make it hard to issue outgoing payments
and withdrawals, while enabling payments between different users of the
exchange).  Consequently, a diverse stake distribution (cf.
Section~\ref{sec:wealth}) is vital to protect against takeovers. 



\ifshort\else
\ifshort
\else
\subsubsection*{Evaluation}
\fi
In (game) theory, Bitcoin's resistance to centralization has been both
supported \cite{kroll2013economics,DBLP:conf/aft/KiayiasS21} and
\ifshort
refuted~\cite{DBLP:journals/cacm/EyalS18},
\else
refuted~\cite{FC:EyaSir14,DBLP:journals/access/LiuKXJW19},
\fi
depending on the economic model assumed for the participants' utilities. In practice,
mining pools have been observed as early as
$2013$~\cite{DBLP:journals/ieeesp/GervaisKCC14}. Between $2016-2020$,
pools created $98.6$\% of Bitcoin
blocks~\cite{DBLP:journals/corr/abs-2002-02082}, with $5$ pools consistently
contributing between $65-85$\% of the eventual blocks and $25$
controlling more than $94$\% of all hashing
\ifshort
power~\cite{measurement2020}.
\else
power~\cite{measurement2020,DBLP:conf/icde/LinLZC21}. 
\fi
Centralization has also been
observed within mining pools. Between $2017-2018$, no entity
controlled more than $21$\%~\cite{FC:GBEvS18} of hashing power, but three pools controlled 
a majority; within these pools, a few participants
($\leq 20$) received over $50$\% of
rewards~\cite{DBLP:journals/corr/abs-1905-05999}. Miners often
participate in multiple pools at the same time, a behavior also observed in
Ethereum~\cite{DBLP:conf/infocom/ZengCCZGXM21}. Although
centralization around pools is high in (PoW-based) Ethereum (in $2019$, $3$
pools controlled a majority of mining
power~\cite{lin2021measuring}), power within the pools is
spread across hundreds of
addresses~\cite{DBLP:conf/infocom/ZengCCZGXM21}, albeit some possibly
owned by the same parties.

\fi

\ifshort\else

Some systems use a committee-based approach, as opposed to Bitcoin's open participation model.
Here, at each time there exists a known designated party which proposes a block and a committee of participants  
that vote for it. The following are examples of such systems, where each
employs their own consensus protocol and defines a different number of
participants per epoch using an on-chain process:\footnote{Sources: 
\href{https://hub.cosmos.network/main/validators/overview.html}{hub.cosmos.network},
\href{https://wiki.polkadot.network/docs/learn-staking}{wiki.polkadot.network},
\href{https://developers.eos.io/welcome/latest/protocol-guides/consensus\_protocol}{developers.eos.io},
\href{https://docs.harmony.one/home/network/validators/definitions/slots-bidding-and-election}{docs.harmony.one},
\href{https://near.org/validators}{near.org}
[August $2022$]}
\begin{inparaenum}[i)]
    \item Cosmos: 175;
    \item Polkadot: 297;
    \item EOS: 21;
    \item Harmony: 800;
    \item NEAR: 100.
\end{inparaenum}
In all these systems, well-known exchanges are among the top elected
validators.\footnote{For example, Binance is a validator in all 
mentioned systems. 
[\href{https://cosmoscan.net/cosmos/validators-stats}{Cosmos},
\href{https://polkadot.subscan.io}{Polkadot},
\href{https://eosauthority.com/producers\_rank}{EOS},
\href{https://staking.harmony.one/validators/mainnet}{Harmony},
\href{https://www.stakingrewards.com/earn/near-protocol}{NEAR}; August
$2022$]}
Interestingly, the stake controlled by the elected validators is mostly
delegated, instead of self-owned. Also, organizations often
control multiple validators, so the number of real actors
is often even smaller than the nominal number of participants
(nevertheless some systems, e.g., Polkadot, go to greater lengths
to ensure the representative participation satisfies desirable properties such
as proportionality, cf. \cite{DBLP:conf/aft/CevallosS21}).
Consequently, identifying the participation distribution among
\emph{real-world users} and the refreshment rate of the
elected committee across multiple epochs is an interesting research question.
Similarly for investigating all the desiderata of representative participation
from a social choice perspective. 
%

\fi

\section{Cryptocurrency Economics}\label{sec:wealth}

A core component of ledger systems is their native token.
Tokens compensate system maintenance and accommodate value transfers. 
They are treated as currency or
assets by their users, thus forming a market economy.
To record data on the
ledger, \eg payments or interactions with
applications, users obtain tokens to pay 
the corresponding fees. System maintainers get compensated in 
tokens to offset their costs. 
Several studies have considered the distribution of tokens and 
their availability (\eg on exchanges) as integral parts of a 
blockchain and its eventual degree of 
decentralization~\cite{DBLP:journals/fbloc/SaiBG21,cheng2021decentralization,FC:RonSha13,moore2013beware}.
Accordingly, in this
section, we explore decentralization in blockchain-based economies, in terms of
initial token distribution, token ownership, and secondary
markets.

\smallskip \noindent \textbf{\emph{Initial Token Distribution.}}
To bootstrap the system, a blockchain protocol defines two parameters:
\begin{inparaenum}[i)]
    \item the distribution of tokens at the system's launch, and
    \item how new tokens are generated and distributed as the system evolves.
\end{inparaenum}
Thus, the generated tokens form the \emph{resource} of interest, while the
\emph{relevant parties} are the token holders.

As with other aspects, Bitcoin led the way and other systems explored alternatives.
In Bitcoin, no coins existed prior to its beginning, \ie there was 
no ``pre-mine.''
Starting from genesis, each block creates a predetermined amount of coins, 
based on a rate that converges to $21$ million tokens in 
existence~\cite{nakamoto2008bitcoin}. 
New coins, along
with transaction fees, are awarded to the miner that
produces each block. Therefore, to acquire new tokens a user 
gathers enough computing power to produce a block. 
In other blockchain systems, some tokens were sold via traditional
markets before the blockchain was deployed. This approach, termed ``Initial Coin
Offering'' (ICO), enabled funding the project with the future
proceeds of the token investment. In return, investors acquired a pre-launch amount of
tokens, which was codified in the
chain's first block.
In terms of token generation, most systems employ a variation of Bitcoin's mechanism, \eg
Ethereum blocks yield $2$ new tokens, while others, like Cardano,
employ elaborate mechanisms to incentivize pooling around a target
number of pools~\cite{DBLP:conf/eurosp/BrunjesKKS20}.

The initial token distribution is particularly important in PoS systems,
where Sybil resilience relies on it (cf. Section~\ref{sec:consensus}). If 
centralized around a
few parties, \eg via pre-mining (or ``pre-minting''), early investors
have to maintain the system in its early stages, while also
receiving the early blocks' rewards. Fewer consensus 
participants during this time lowers the threshold for adversarial 
takeover, threatening the system's \emph{safety} and 
\emph{liveness}. In both PoW and 
PoS systems, new users are onboarded 
if early investors sell tokens 
on secondary markets. Consequently, early investors
control the system's expansion and valuation, impacting its 
\emph{stability}.

Finally, the process of distributing tokens might be elemental for
privacy-oriented systems. Typically, such projects employ zero-knowledge
protocols that rely on a secure construction of a common reference string
(CRS). If the CRS's construction is centralized, then the party that creates it
can deanonymize all transactions or violate their correctness. 
To avoid such hazards, various ceremony
protocols have been proposed in order to construct the CRS in a distributed
manner~\cite{zcash-ceremony,DBLP:conf/fc/KerberKK21,nikolaenko2022powers,AC:KMSV21}.

\ifshort\else
\ifshort
\else
\subsubsection*{Evaluation}
\fi
PoS systems like Cardano, NEO, and Algorand tried to reduce early-stage risks
via a two-phase launch. At first, the ledger was  
controlled by either the core development company or
foundation or a committee numbering a small number of
entities. After token ownership was sufficiently distributed,
participation opened widely to all stakeholders.
Beyond the obvious issues in maintaining a
permissioned database, the first phase typically takes years to conclude.
Early users often tend to either not
participate or transfer their tokens to the few exchanges
that support these new tokens~\cite{victor2021taxonomy}. Therefore, an interesting
question is the relationship of the delay between launch and full
decentralization and the diversity
of early investors.
%

\fi

\smallskip \noindent \textbf{\emph{Token Ownership.}}
Diverse token ownership plays a central role in the usability and security of a
blockchain. Hence, the system's circulating tokens 
are the \emph{resource} of interest, while the \emph{relevant 
parties} are:  
\begin{inparaenum}[i)]
    \item key managers, and
    \item legal asset owners.
\end{inparaenum}
This distinction arises due to the existence of custodians, who control assets
on behalf of other stakeholders, and users controlling multiple
addresses.

If most tokens are owned by a few parties, many hazards arise.
First, PoS systems' security, \ie \emph{safety} and \emph{liveness}, relies
directly on diverse token ownership,
which makes corrupting enough parties to control a majority of tokens more
threatening. 
Second, the
token's price may be manipulated, posing a risk on the system's
\emph{stability} and, indirectly, security, in both PoS and PoW systems. Specifically, participation
cost, \eg for mining equipment or electricity, is denominated in fiat currency.
However, miner income from block rewards comes
in tokens. Thus, miners need to sell part of the rewards (for fiat) to
pay for their operational costs.
If the market is volatile, profitability is more precarious and miners are
possibly less inclined to participate, which can impact the \emph{safety} or
\emph{liveness} of the system by reducing the threshold for conducting a 51\%
attack.

Various factors drive token ownership centralization. Initial
tokens are often allocated centrally (see above). System incentives, \eg fixed
token supply, generally favor hoarding tokens instead of spending them.  
Finally,
rich participants may accumulate capital faster than small
ones, an inevitability in pseudonymous
systems where downwards wealth redistribution is 
impossible~\cite{karakostas2019cryptocurrency}.

\ifshort\else
\ifshort
\else
\subsubsection*{Evaluation}
\fi
Bitcoin's wealth ownership and transaction graph has been
analyzed since at least $2012$~\cite{FC:RonSha13}. Over time, it demonstrated a
three-phase history of distinct (de)centralization patterns, where $100$
addresses possess a high centralization degree of assets and wealth flow in the
network~\cite{cheng2021decentralization,DBLP:journals/fbloc/SaiBG21}. Similar
analyses exist for Ethereum~\cite{chen2020understanding},
Zcash~\cite{USENIX:KYMM18}, and other
cryptocurrencies~\cite{motamed2019quantitative}. 

As of $2022$, cryptocurrency wealth concentration is particularly
extreme (Table~\ref{tab:gini}). To establish some context, the income Gini coefficient of the $10$
lowest-performing countries ranges between $0.63-0.512$~\cite{world-bank-gini}.
Bitcoin has a Gini coefficient of $0.514$,
considering only the $10,000$ richest addresses, and a staggering $0.955$ \wrt
all addresses. In the arguably deeply unequal global
real-world economy, the richest $0.01$\% of individuals ($520,000$ people) hold
$11$\% of all wealth~\cite{fortune-billionaires}. Bitcoin
manages to beat that figure, with $100$ addresses
holding $14.01$\% of all tokens.


\begin{table}[ht]
    \centering \def\arraystretch{1.5}

    \begin{center}
      \footnotesize
        \begin{tabular}{|c|c|c|c|c|}
            \hline
               \textbf{System}
             & \textbf{Addresses}
             & \textbf{Top-100}
             & \textbf{Gini (10K)}
             & \textbf{Gini} \\
             \hline

             Bitcoin & 42,943,534 & 14.01 \% & 0.5145 & 0.956 \\
             \hline

             Ethereum & 193,673,067 & 39.75 \% & 0.6757 & 0.978 \\
             \hline

             Dogecoin & 4,839,762 & 68.49 \% & 0.8297 & 0.986 \\
             \hline

             Zcash & 345,766 & 32.92 \% & 0.7796 & 0.974 \\
             \hline

             Bitcoin Cash & 17,123,166 & 28.4 \% & 0.672 & 0.97 \\
             \hline

             Litecoin & 5,442,751 & 36.8 \% & 0.6757 & 0.978 \\
             \hline

             Ethereum Classic & 511,491 & 41.37 \% & 0.8289 & 0.988 \\
             \hline

             \ifshort
             \else
             Dash & 1,545,061 & 14.68 \% & 0.6896 & 0.971 \\
             \hline
             \fi
        \end{tabular}
      \normalsize
    \end{center}
    \caption{
        Cryptocurrency wealth distribution:
        i) addresses that control assets; 
        ii) percentage of wealth controlled by the top 100 wealthiest
        addresses; 
        iii, iv) Gini coefficient of 10K wealthiest and all addresses
        resp. (Sources:
        \href{https://github.com/blockchain-etl/public-datasets}{Blockchain ETL},
        \href{https://cloud.google.com/bigquery}{Google BigQuery},
        \href{https://www.coincarp.com/}{CoinCarp}. [April 2022])
    }
    \label{tab:gini}
\end{table}


A complexity in measuring wealth decentralization in
cryptocurrencies arises due to their pseudonymous (or even anonymous) nature. Specifically,
the number of addresses often does not correspond to individual people or entities,
cf. \cite{FC:RonSha13,DBLP:journals/cacm/MeiklejohnPJLMV16}.
A user may control multiple addresses, \eg each with a small
balance. When interpreting the Gini coefficient, this artificially enlarges the population and possibly biases the results
towards decentralization. In addition, an address's assets may be owned by
many users (\eg exchange addresses), which
biases Gini towards centralization. Thus, developing tools to
compute wealth inequality in blockchain systems, without sacrificing
core features like anonymity and privacy, is a crucial problem for
exploration.
%

\fi

\smallskip \noindent \textbf{\emph{Secondary 
Markets.}}\label{subsec:secondary-markets}
Distributing the tokens to a wide population is predominantly made on secondary
markets. The rate of token production is typically slow, depending on block
production, and the new tokens are often distributed to existing users. 
Therefore, new users are onboarded via centralized exchanges and,
to a lesser extent, face-to-face transactions.
The tokens that are 
bought and sold through these markets constitute the \emph{resource} of 
interest, when it comes to measuring the decentralization of secondary markets, 
while the \emph{relevant parties} are 
\begin{inparaenum}[i)]
    \item the assets for which they are bought and sold (``trading 
    pairs''),\footnote{In our context the liquidity of a trading pair 
    is measured across \textit{all} exchanges.}, and
    \item the exchanges that host these trades.
\end{inparaenum}

Many hazards arise when tokens are available on limited markets.
First, exchanges offer little \emph{privacy} guarantees, so their
operators have full access of user data, following KYC regulations. 
Second, exchanges are largely \emph{unregulated} by financial authorities and 
may engage in market manipulation. 
Third, few marketplaces often result in lower liquidity. Thus, the threshold 
for manipulating the token's price by some percentage, via selling or buying 
tokens, also lowers. 
Similarly, if most of the token's liquidity is allocated to a few trading 
pairs, then it becomes exposed to the problems of the tokens at the other end 
of the pairs (\eg the collapse of one of these systems might trigger a huge 
liquidity loss). 
All such events threaten the system's \emph{stability}, while also, when mining 
profitability drops due to the token's devaluation, \emph{safety} and 
\emph{liveness} are indirectly hurt.

\ifshort\else
\ifshort
\else
\subsubsection*{Evaluation}
\fi
Table~\ref{tab:secondary-markets} summarizes secondary blockchain market data 
across $121$ exchanges. Many systems (Bitcoin, Ethereum,
Litecoin, XRP) are traded on all but a few small exchanges. Tether is by far the most
available, in terms of market pairs, and used, in terms of volume.
Interestingly, for all systems, except perhaps Bitcoin, the majority of volume
is not of the highest transparency.\footnote{For ``transparency'' see the
methodology and data of Nomics:
\url{https://nomics.com/blog/essays/transparency-ratings}.} This is consistent
with reports that show market manipulation is endemic in cryptocurrency
markets, with multiple cases of wash trading, fake trading volumes, 
and other fraudulent
behavior~\cite{coinbase2021sec,bti2020report,bitwise2019report}.
Market transactions are primarily conducted in a handful of
exchanges. By far the most used is Binance ($20$\% of the total daily
volume),\footnote{\href{https://coinmarketcap.com}{CoinMarketCap} [August 2022]} although
Coinbase is the most recognized in North America~\cite{nomics2021review}.


\begin{table}[ht]
    \centering \def\arraystretch{1.5}

    \begin{center}
      \footnotesize
        \begin{tabular}{|c|c|c|c|c|}
            \hline
               \textbf{System}
             & \textbf{Exchanges}
             & \textbf{Pairs}
             & \textbf{Transparent Volume} \\
             \hline

             Tether & 80 & 21691 & \$$52.86$B ($44$\%) \\
             \hline

             Bitcoin & 100 & 9622 & \$$41.31$B ($56$\%) \\
             \hline

             Ethereum & 103 & 5680 & \$$14.24$B ($42$\%) \\
             \hline

             Dogecoin & 84 & 1856 & \$$175.5$M ($29$\%) \\
             \hline

             \ifshort
             \else
             XRP & 90 & 555 & \$$551.8$M ($26$\%) \\
             \hline

             Bitcoin Cash & 90 & 537 & \$$80$M ($9$\%) \\
             \hline

             Litecoin & 98 & 485 & \$$210$M ($25$\%) \\
             \hline
             \fi

             Cardano & 78 & 437 & \$$1.03$B ($41$\%) \\
             \hline

             Solana & 75 & 383 & \$$1.25$B ($39$\%) \\
             \hline

             \ifshort
             \else
             Polkadot & 77 & 378 & \$$195.26$M ($36$\%) \\
             \hline

             Ethereum Classic & 67 & 350 & \$$34.7$M ($14$\%) \\
             \hline

             Zcash & 59 & 320 & \$$38.6$M ($20$\%) \\
             \hline
             \fi

             Avalanche & 64 & 318 & \$$574.29$M ($49$\%) \\
             \hline

             Monero & 41 & 101 & \$$39.9$M ($22$\%) \\
             \hline
        \end{tabular}
      \normalsize
    \end{center}
    \caption{
        Secondary market data across $121$ exchanges (June $2022$):
        i) exchanges and trading pairs
        (\href{https://coinmarketcap.com/}{\emph{CoinMarketCap}});
        ii) transaction volume rated as ``transparent'' by
        \href{https://nomics.com}{\emph{nomics}}.
    }
    \label{tab:secondary-markets}
\end{table}


\fi

\section{Client API}\label{sec:clients}

To join a blockchain system, full nodes need to download and
parse the entire ledger, which often amounts to hundreds of
GBs.\footnote{Bitcoin: $485$ GBs;
Ethereum: $819$ GBs. [\href{https://bitinfocharts.com/bitcoin}{bitinfocharts},
\href{https://etherscan.io/chartsync/chaindefault}{Etherscan}; August
$2022$]} The ledger's state, which is usually stored in memory,
is also large\footnote{Bitcoin's UTxO set is
$4.71$ GBs.
[\href{https://statoshi.info/d/000000009/unspent-transaction-output-set?orgId=1&refresh=10m}{Satoshi
info}; August $2022$]} and often poorly 
maintained~\cite{EPRINT:KarKarKia21}.
Consequently, maintaining a full node requires significant computational and
storage capacity and, eventually, becomes impossible to host on home equipment.

This concern is well-known and ongoing research tries to resolve it via ledger
compression~\cite{EPRINT:KiaMilZin17,EPRINT:BMRS20a,EPRINT:KatBon20}. In
practice though, users often employ third-party services that offer an
interface to the ledger~\cite{marlinspike2022my}. Given the widespread use and
variety of applications that rely on such services, the ``client API'' layer
can be susceptible to
centralization~\cite{chatzigiannis2022sok}. The \emph{resources} we identify in
this case are tokens, which are stored in wallets without ledger verification
capabilities, and the \emph{relevant parties} the full node operators that
service them. We note that solutions like Simplified Payment Verification
(SPV)~\cite{nakamoto2008bitcoin} or succinct verification
proofs~\cite{EPRINT:DKKZ20,EPRINT:KiaPolZin20,EPRINT:KiaMilZin17,SP:BKLZ20},
which are not full nodes but do validate the ledger to some extent, are not
considered here. Instead, we focus on wallets that rely entirely on a
trusted node for access to the ledger's content.

Many properties are at risk for such wallets by corrupted full node services.
For example, the service could perform a double-spending attack (\emph{safety}
violation), by presenting to the wallet a transaction that is not in fact
published on the ledger; observe that, without any access to the ledger itself,
the wallet needs to trust the data presented by the node.  Similarly, since the
wallet relies on the full node for transaction processing and balance
computations, \emph{liveness}, \emph{privacy}, and \emph{stability} hazards
arise, as the node can block, de-anonymize or, depending on 
the implementation,
divert a user's funds and transactions.

\ifshort\else
\ifshort
\else
\subsubsection*{Evaluation}
\fi
In Bitcoin, most wallets are either SPV or
explorer-based~\cite{karantias2020sok}. In the first case, the wallet obtains
the chain's headers and, to verify that a transaction is published,
requests a proof from full nodes. Although SPV does mitigate safety attacks, it
also hurts the user's privacy, as their transaction information is leaked to
the full node operators. Explorer-based wallets instead rely entirely on a 
single explorer service and its full nodes, which are trusted completely.
In $2018$, $5-10$\% of all Ethereum nodes reportedly relied on a
centralized blockchain API service, Infura~\cite{infura-usage}. This reliance
continued throughout the years. In $2020$, a service outage demonstrated in
practice the hazards of such centralization~\cite{infura-outage}. In $2022$, a
misconfiguration on Infura's part resulted in wallets (and, thus, user funds)
being inaccessible~\cite{infura-bug}.  In terms of applications, OpenSea is the
leading hosting service for NFTs. As of $2021$, it reportedly handled $98$\% of
all NFT volume~\cite{opensea-usage}, charging a $2.5$\% commission on all
sales. As expected, an OpenSea outage in $2022$ also resulted in the NFT
market being practically unusable~\cite{opensea-outage}.

\fi

\section{Governance}\label{sec:governance}

Governance in blockchain systems is a broad
topic~\cite{DBLP:journals/corr/abs-2201-07188,pelt2021defining,beck2018governance}
 and of high interest among studies of blockchain 
decentralization~\cite{DBLP:journals/ieeesp/GervaisKCC14,FCW:AzoMalMei18,DBLP:journals/ipm/SaiBFG21}.
Here, we focus on two aspects:
\begin{inparaenum}[i)]
    \item improvements and conflict resolution;
    \item fund allocation for research and development (R\&D).
\end{inparaenum}

\smallskip \noindent \textbf{\emph{Improvements and Conflict 
Resolution.}}
Decision-making mainly concerns conflicts that arise regarding 
potential blockchain modifications
and improvement proposals. Proposals 
may affect \emph{mining}, \eg
changing the PoW function~\cite{monero-change} or switching to
PoS~\cite{ethereum-merge}, the \emph{consensus protocol}, \eg changing block
structure~\cite{bitcoin-cash-fork}, or \emph{token ownership}, \eg denylisting~\cite{ethereum-fork}. 
In theory, anyone can propose changes in blockchain systems and respond in 
some way, depending on their role. 
In essence, full nodes assume executive, legislative, and judicial powers
by operating the ledger and choosing its rules,
while other actors voice their opinions by affecting the token's market
price~\cite{alston2019constitutions}. The governance
\emph{resource} is decision-making power, which
may take various forms, and the \emph{relevant parties} are all active
entities in the system.

If the other relevant layers are centralized, governance follows suit.
For instance, if mining is concentrated around a few operators, they might 
force a choice by mining on one ledger.
Where a voting mechanism is employed to reach a decision, voting power 
typically corresponds to each participant's wealth, with each token granted one 
vote (due to the pseudonymous nature of these systems).
If ownership is concentrated around a small number of stakeholders, the 
decisions might aim at benefiting these few parties in the short term, at the 
expense of the system's long-term benefit.
If a disagreement turns into a stalemate, systems may split into distinct
ledgers, that share the same history up to a point but diverge
thereupon~\cite{bitcoin-forks,ethereum-fork}. These outcomes harm
the system's \emph{stability}, and indirectly 
threaten its \emph{safety} and \emph{liveness}. 

An effective governance process should prevent such harmful events.
However, it is not always possible to make
sound decisions in a decentralized manner, as demonstrated by theoretical
results in social choice such as Arrow's  
impossibility theorem~\cite{arrow1950difficulty}.\ifshort~\else\footnote{The 
theorem's formal
proof states that no rank-order electoral system can satisfy a set of three
criteria:
i) if every voter prefers X over Y, the group prefers X over Y;
ii) if every voter's preference between X and Y remains unchanged, the group's
preference also remains unchanged (even if voter preferences over other pairs
change);
iii) there is no ``dictator'', \ie no single party can always determine the
group's preference. 
This idea dates back to the late $18$th century and Condorcet's paradox in
collective decision-making: consider three candidates $A, B, C$ and three
voters $V_1, V_2, V_3$, with (ordered) candidate preferences $[A, B, C], [B, C,
A], [C, A, B]$ resp; although each candidate's order of preference is
consistent, the collective preference is cyclical.}\fi
Additionally, when agents act in a selfish manner, as is presumed in distributed
ledgers, efficiency can degrade (cf. ``Price of
Anarchy''~\cite{koutsoupias1999worst}).
Therefore, decentralized decision-making processes face a challenge, as they 
need to address various social choice theory (\eg Arrow's theorem) and 
game-theoretic (\eg rational ignorance~\cite{downs1957economic}) considerations.

\ifshort\else
\ifshort
\else
\subsubsection*{Evaluation}
\fi
Most systems employ an \emph{Improvement Proposal} mechanism,
where proposals are posed as issues in Github, a (centralized) system that is extensively used for software development. If a change gathers enough support,
it is incorporated in the codebase. To voice approval for proposals, miners
often include encoded messages in blocks.
From early on, proposals in Bitcoin and Ethereum have been made by a
handful of developers~\cite{DBLP:journals/ieeesp/GervaisKCC14,FCW:AzoMalMei18}.
In the discussion phase, many people participate but again only a
few actors contribute most comments, while in cases like
Bitcoin the groups of developers and commenters largely overlap.

\fi

\smallskip \noindent \textbf{\emph{R\&D Funding.}}
Funding for research and  development can cover the maintenance of legacy 
codebase, research in features like privacy and scalability, market incentives, 
\eg stabilizing the token's price at times of high volatility, and more. Thus, 
the \emph{resource} of interest here is capital and the \emph{relevant parties} 
are the active researchers and developers in a ledger's ecosystem.

Ledgers typically make no funding provisions, besides allocating
rewards from coin issuance and transaction fees. R\&D is conducted
via corporate vehicles which rely on traditional funding models, such as venture
capital. However, since designing and implementing hardware
and software for distributed ledgers is particularly expensive,
this model can lead to centralization, as
discussed in Sections~\ref{sec:hardware} and~\ref{sec:software}. In addition,
lack of funding or concentration around a few teams may delay crucial
updates or new features, thus hindering \emph{stability}.

A common alternative to traditional financing is ledger ``self-funding.'' 
Here, the system defines a \emph{treasury}, \ie a pot which accumulates funds
over time
that are allocated for R\&D~\cite{allen2021trust,NDSS:ZhaOliBal19}. A treasury
is typically managed collectively by the ecosystem, often via an open and
inclusive process where anyone can submit proposals and the system's
stakeholders vote for funding allocation. Therefore, a treasury can help
nurture a diverse ecosystem of development teams, albeit it is
not, on its own, a sufficient condition for decentralized R\&D funding.

\ifshort\else
\ifshort
\else
\subsubsection*{Evaluation}
\fi
Most existing blockchain systems follow the first approach, \ie not making
funding provisions. In many cases,
funding is channeled through a few foundations and
companies.\footnote{The first usually take the name of the token, \eg the
\{Bitcoin, Ethereum, Cardano\} Foundations. Examples of the second are the ASIC
companies discussed in Section~\ref{subsec:physical-hardware} or software 
companies
like Blockstream (Bitcoin), Consensys (Ethereum), Input Output (Cardano), \etc}
Treasuries are present in some ledgers, like Decred, Cardano, and Dash.
Despite their potential though, widespread funding has yet to be 
demonstrated for most systems.\footnote{Decred's treasury holds
\$$23.8$M, and has allocated $\$250$K over the past year. Cardano's
treasury holds approx. \$$500$M and has distributed \$$17.2$M across $939$
projects. Dash, one of the first systems to set a treasury, allocated
\$$500,000$ over $2018$, but it appears non-functional as of
$2022$. 
[\href{https://dcrdata.decred.org/treasury}{dcrdata.decred.org},
\href{https://bit.ly/FundedProjectsReporting}{cardano.ideascale.com},
\href{https://dashvotetracker.com}{dashvotetracker}; August $2022$]}

\fi

\section{Geography}\label{sec:geography}

Geographic decentralization is a key point of interest 
\cite{mariem2020all,sun2022spatial,DBLP:journals/ipm/SaiBFG21}, 
and it touches upon all layers covered in the
previous sections. Accordingly, it involves all resources described so far,
\eg hashing power or tokens. Nonetheless, it constitutes a
dimension on its own, as parts of a system may be well distributed \wrt one
dimension but geographically concentrated.\footnote{For example, independent
actors may participate in mining within a single
country.}

The tendency to centralize in certain areas
arises due to economical, technological, or sociopolitical factors. For
example, miners often set up their operations in countries with
low electricity costs, hardware companies operate in countries with small
production costs, nodes are hosted in areas with high internet speed, and tokens
are accumulated by residents of countries with low taxes and where many
exchanges operate. Geographical centralization poses two main threats to the
properties of a ledger: 
\begin{inparaenum}[i)]
    \item physical hazards and
    \item legal impediments.
\end{inparaenum} 

\smallskip \noindent \textbf{\emph{Physical safety.}}
Physical hazards 
could threaten a system's infrastructure. If part of a system is located
in a small area, connectivity failures or outages could destroy or split the ledger's network.\footnote{Although exogenous hazards, like natural disasters, are outside the scope of this work, concentration in a small area can enable weaker adversaries to disrupt a ledger's execution. For example, an adversary could isolate a building or a single computing center from the rest of the network, although assuming an adversarial disruption of the grid of an entire country, or continent, can be considered unrealistic.}
This
concern is particularly relevant in PoW, where equipment is hard to relocate. 
All \emph{resources} examined above
can be impacted (\eg 
via drops in hashing power or token loss when mining equipment or cold storage 
is damaged), while the 
\emph{relevant parties} are the regions of
resource concentration. 
Single points of failure may arise when
geographically-concentrated nodes act as central hubs, harming
either \emph{safety} or \emph{liveness} (cf. Section~\ref{sec:p2p}) and,
indirectly, \emph{stability} (\eg due to increased market volatility).

\smallskip \noindent \textbf{\emph{Legal compliance.}}
Failures can also possibly occur due to legal pressures.
If some layer is concentrated in a specific
jurisdiction, authorities can possibly restrict or
subvert it. Again, this touches upon all \emph{resources} examined so far, as 
all are
influenced by the law, with the \emph{relevant parties} being legal
jurisdictions. Depending on the occasion, different properties of the system
are impacted. For example, if a country bans  Bitcoin mining, the power drop could decrease
the threshold for controlling a majority (\emph{safety} hazard), while 
blocks are produced at a slower pace until the PoW difficulty is
recalculated (\emph{liveness} hazard). \emph{Stability} could also be hurt, if
some part of the system, \eg mining, software access, or asset
ownership is restricted. Additionally,
exchanges can be legally bound to follow KYC procedures to comply with AML
regulations~\cite{amlEnforcement}, linking the users' identities to their
activity something that may lead to compromising their \emph{privacy}. Arguably, a system is
more likely to uphold its properties by falling under many
jurisdictions, such that violating the properties
requires the coordinated
efforts of multiple authorities.

\ifshort\else
\ifshort
\else
\subsubsection*{Evaluation}
\fi
In $2014$, $37$\%
of Bitcoin nodes resided in the US and China~\cite{DBLP:conf/fc/DonetPH14}.
\ifshort
\else
In $2018$, its testnet also showed a concentration in the USA, Central Europe, 
and East Asia~\cite{FC:DBPLPMB19}. 
\fi
In $2019$, Bitcoin mining
hardware was mostly located in China (particularly Sichuan) and the US~\cite{bendiksen2019bitcoin,kohler2019life}.
Notably, the mining pools then-located in China accounted for
$68$\% of all hashrate~\cite{stoll2019carbon}.
As of $2022$, a large fraction of nodes communicates over
Tor,\footnote{$46.5$\% of Bitcoin's nodes operate over Tor. 
[\href{https://bitnodes.io/dashboard/}{bitnodes}; August $2022$]}, thus
analyzing the network's topology is often hard. 
Nonetheless,
more than $\frac{1}{3}$ of Bitcoin mining is presumably located in the USA, with Kazakhstan and
Russia following with $18$\% and $11$\% respectively~\cite{cbecimining}.
In terms of full nodes, USA and Germany
see roughly equivalent shares, with other countries hosting far fewer;
still, a majority communicates over Tor.
Until $2021$ China hosted as high as $70.9$\%~\cite{cbecimining} of Bitcoin
mining power; following its ban that year, Bitcoin's hashrate dropped from $197$ to
$68$ Ehash/s in one
month.\footnote{\href{https://bitinfocharts.com/comparison/bitcoin-hashrate.html}{bitinfocharts.com}}
Ethereum (pre PoS) observed similar concentration
patterns; by far the most nodes are located in the USA
($37$\%) and, secondarily, Germany ($16.66$\%)~\cite{chainstack}.
Finally, Monero nodes are mostly located in the US and, to a lesser extent,
elsewhere~\cite{FC:CYDLV20}. 
Table~\ref{tab:countries} shows various systems' geographical distribution.


\begin{table}[ht]
    \centering \def\arraystretch{1.5}
    
    \begin{center}
        \footnotesize
        \begin{tabular}{|c|c|c|c|c|c|}
            \hline
            
            \multirow{5}{*}{ \begin{tabular}[c]{@{}c@{}} \textbf{Bitcoin} \\ (\emph{Total: $7670$})\end{tabular} } & US & $2862$ & 
            \multirow{5}{*}{ \begin{tabular}[c]{@{}c@{}} \textbf{Ethereum} \\ (\emph{Total: $5884$})\end{tabular} } & US & $2781$ \\
            \cline{2-3}\cline{5-6}                       & DE & $1164$ &
            & DE & $678$ \\
            \cline{2-3}\cline{5-6}                       & FR & $505$ &
            & SG & $283$ \\
            \cline{2-3}\cline{5-6}                       & CA & $377$ &
            & UK & $241$ \\
            \cline{2-3}\cline{5-6}                       & NL & $309$ &
            & FI & $236$ \\
            \hline
            
            \multirow{5}{*}{ \begin{tabular}[c]{@{}c@{}} \textbf{Bitcoin Cash} \\ (\emph{Total: $987$})\end{tabular} } & US & $419$ &
            \multirow{5}{*}{ \begin{tabular}[c]{@{}c@{}} \textbf{Dogecoin} \\ (\emph{Total: $1096$})\end{tabular} } & US & $475$ \\
            \cline{2-3}\cline{5-6}                            & DE & $148$ &
            & DE & $187$ \\
            \cline{2-3}\cline{5-6}                            & FR & $69$ &
            & FR & $71$ \\
            \cline{2-3}\cline{5-6}                            & NL & $35$ &
            & CA & $47$ \\
            \cline{2-3}\cline{5-6}                            & CA & $35$ &
            & CN & $33$ \\
            \hline
            
            \ifshort
            \else
            \multirow{5}{*}{ \begin{tabular}[c]{@{}c@{}} \textbf{Zcash} \\ (\emph{Total: $9056$})\end{tabular} } & US & $2913$ &
            \multirow{5}{*}{ \begin{tabular}[c]{@{}c@{}} \textbf{Litecoin} \\ (\emph{Total: $1182$})\end{tabular} } & US & $330$ \\
            \cline{2-3}\cline{5-6}                     & DE & $2539$ &
            & DE & $221$ \\
            \cline{2-3}\cline{5-6}                     & FR & $1164$ &
            & FR & $149$ \\
            \cline{2-3}\cline{5-6}                     & CA & $678$ &
            & CA & $58$ \\
            \cline{2-3}\cline{5-6}                     & PL & $298$ &
            & RU & $54$ \\
            \hline
            \fi
        \end{tabular}
        \normalsize
    \end{center}
    \caption{
        Geographical distribution of full nodes in
        various ledger systems. (\href{https://blockchair.com/}{blockchair},
        \href{https://ethernodes.org/}{ethernodes}; June $2022$) 
    }\label{tab:countries}
\end{table}


In terms of legal jurisdiction, different aspects are
centralized in different countries. 
In Bitcoin, the $4$ companies that predominantly produce mining hardware\footnote{\href{https://www.bitmain.com/}{Bitmain},
\href{https://www.microbt.com/}{MicroBT}, \href{https://canaan.io/}{Canaan},
\href{https://www.ebang.com.cn/}{Ebang}.} are all based in China. 
Regarding secondary markets, many exchanges operate in multiple countries (Table~\ref{tab:secondary-markets});
$20$ of $121$ operate in USA, thus falling under US jurisdiction, $17$ in
China, and $10$ in Japan, with the rest spread across the world.
However, only $8$ are based in the US, with most registered in the
Seychelles ($13$) and other ``offshore'' locations. Many ICOs
also exclude US investors, following their US classification as
securities~\cite{securities2017sec}. 
Finally, an interesting   case concerns the Bitcoin Core software, which is not
available via \href{https://bitcoin.org}{bitcoin.org} in the UK, following a
related court ruling~\cite{ukBitcoinCore}. 

\fi

\section{Case Studies}\label{sec:case-studies}

We now apply our methodology to a number of case studies. First,
we review Bitcoin's status \wrt each identified layer, 
showcasing how a project can be analyzed across all strata. 
Second, we apply the Minimum Decentralization Test (MDT) (cf. 
Definition~\ref{def:mdt}) on an array of projects and show that 
they fall short due to centralization some layer --- 
specifically the governance layer.

\smallskip \noindent \textbf{\emph{Cross-layer Study: 
Bitcoin}}\label{sec:bitcoin}

\ifshort
\noindent \emph{Hardware.}
\else
\subsubsection*{Hardware}
\fi
No concrete data could be found on the distribution of hashing power across PoW
mining products. Although hundreds of ASICs are available, on top of
generic hardware (\eg GPUs), as of $2022$ the market appears centralized
around $4$ ASIC manufacturers~\cite{tokeninsight-2}. Interestingly, only some
ASICs are profitable, so, unless the token's price increases without an
increase in PoW difficulty, mining should be expected to concentrate around
these products.\footnote{A profitability calculator is available at
\href{https://www.nicehash.com/profitability-calculator}{nicehash.com}.}

\ifshort
\noindent \emph{Software.}
\else
\subsubsection*{Software}
\fi
An overwhelming majority ($99.15$\%) of full nodes run Bitcoin
Core,\footnote{Source:
\href{https://blockchair.com/bitcoin/nodes}{blockchair.com}}
so Bitcoin is completely centralized around (releases of) 
this product. 
Regarding the distribution of tokens across wallets, no data could be found.

\ifshort
\noindent \emph{Network.}
\else
\subsubsection*{Network}
\fi
Regarding peer discovery, Bitcoin Core sets $8$ outgoing and $125$ incoming
connections, chosen randomly from known and/or hardcoded peers.
Most Bitcoin nodes communicate over Tor, making topology analyses particularly
hard.\footnote{$53.3$\% of Bitcoin's nodes operate over Tor.
[\href{https://bitnodes.io/dashboard/}{bitnodes}; October $2022$]} Nonetheless,
it is estimated that the network is evenly spread across multiple Autonomous
Systems, thus presenting high levels of decentralization~\cite{SP:ApoZohVan17}.

\ifshort
\noindent \emph{Consensus.}
\else
\subsubsection*{Consensus}
\fi
On the consensus layer, Bitcoin presents mixed results regarding
decentralization (cf. Section~\ref{sec:consensus}). Hashing power is distributed across
thousands of machines. Although no concrete data could be
found, folklore evidence suggests that these machines are owned by a highly
diverse set of users. However, Bitcoin also observes high levels of
centralization around pools, \ie \wrt block formation and the input to the PoW
module; specifically, at the time of writing, 
$4$ pools control more than $75$\% of the whole network's
mining power.\footnote{Source:
\href{https://www.statista.com/statistics/731416/market-share-of-mining-pools/}{statista.com}}

\ifshort
\noindent \emph{Tokenomics.}
\else
\subsubsection*{Tokenomics}
\fi
At its onset, no Bitcoin tokens existed. They were generated and allocated as
the system progressed. Early participants were disproportionately favored, as
half of all tokens were created within the first two years, when consensus
participation was sparse and mining was conducted by only a few parties. As
more transactions were issued, the tokens were distributed more widely, albeit
wealth is still highly centralized, compared to real-world economies (cf.
Table~\ref{tab:gini}). Specifically, approx. $43$M addresses own some amount of
tokens, with the top $100$ addresses controlling $14.01$\% of all wealth.
Nonetheless, tokens are traded on more than $100$ marketplaces at volumes of
approx. \$$53$B (cf. Table~\ref{tab:secondary-markets}).

\ifshort
\noindent \emph{Client API.}
\else
\subsubsection*{Client API}
\fi
Most of the available Bitcoin wallet software is either SPV or
explorer-based~\cite{karantias2020sok}. In the first case, the wallet downloads
only the block headers, so it  does not validate each block's transactions, while
in the second case the wallet relies entirely on a server. However, no data
could be found on the ownership of Bitcoin tokens \wrt wallet types, therefore
Bitcoin's decentralization \wrt the client API layer is inconclusive.

\ifshort
\noindent \emph{Governance.}
\else
\subsubsection*{Governance}
\fi
Deciding on improvement proposals and conflict resolution in Bitcoin is
somewhat centralized, but not entirely. Specifically, decisions, which are made
by accepting suggestions via GitHub, are typically taken by a small set of
developers, who are often the ones to comment during the relevant
discussions~\cite{DBLP:journals/ieeesp/GervaisKCC14,FCW:AzoMalMei18}. 
In terms of development funding, Bitcoin makes no provisions. Therefore, the
available data are inconclusive on how many sources of funding exist, \eg
companies and foundations, and how much influence each has.

\ifshort
\noindent \emph{Geography.}
\else
\subsubsection*{Geography}
\fi
As mentioned earlier, most Bitcoin nodes communicate over Tor, which makes
analyzing the network's topology difficult. Nonetheless, Bitcoin
miners, although fairly well distributed with a presence in $95$ different 
countries, tend to cluster in certain areas.
At the time of writing, more
than $\frac{1}{3}$ of mining is located in the USA, with Kazakhstan and
Russia following with $18$\% and $11$\% respectively~\cite{cbecimining}.
In terms of full nodes (which may not participate in mining), USA and Germany
see roughly equivalent shares, with other countries hosting far fewer nodes 
(cf. Table \ref{tab:countries}), although still a majority communicates 
anonymously.

\medskip \noindent \textbf{\emph{MDT 
Studies.}}\label{sec:mdt-studies}
We now turn our attention to projects that fail the Minimum 
Decentralization Test (MDT), showcasing how the MDT 
can be used to identify points of centralization in blockchain 
systems.

\noindent \emph{Fiat-backed stablecoins.}
A prime example of projects that fail the MDT is fiat-backed stablecoins.
Briefly, in a USD-backed stablecoin system, for each token that is live on the 
ledger there exists \$$1$ which is held in escrow in a company's bank account, 
s.t. at any point in time, a token holder can exchange their tokens for the 
equivalent number of USD. Therefore, the main selling point is that the token 
should always (in theory) be valued by the market at \$$1$. Such projects 
include Tether (USDT), USDC, Gemini Dollar (GUSD), TrueUSD (TUSD), and Binance 
USD (BUSD).\footnote{USDT: \href{https://tether.to}{tether.to},
USDC: \href{https://www.circle.com/en/usdc}{circle.com},
GUSD: \href{https://www.gemini.com/dollar}{gemini.com},
TUSD: \href{https://www.tusd.io}{tusd.io}, 
BUSD: \href{https://www.binance.com/en/busd}{binance.com}} 
In all these systems, there exists a single legal entity which is responsible 
for issuing tokens when receiving USD and redeeming tokens in exchange for the 
USD held in escrow.
Therefore, these entities are single points of control within the
\emph{governance}
layer of each system.

\noindent \emph{Wrapped tokens.}
Another family of systems for which the MDT often fails is bridges. A 
bridge enables transferring assets of one ledger to another, \eg Bitcoin to
Ethereum. This is achieved by creating a ``wrapped'' version of the original
token on the destination, with each wrapped token corresponding to a (frozen)
token on the source side. These systems can also be centralized in the 
governance layer, as a single custodian is typically 
responsible for the creation and destruction of the wrapped tokens. Such bridge 
examples include Wrapped Bitcoin (WBTC),
with BitGo being solely responsible for minting new tokens, and Huobi Bitcoin
(HBTC), with Huobi being the
custodian.\footnote{WBTC: \href{https://wbtc.network}{wbtc.network},
HBTC: \href{https://www.htokens.finance}{htokens.finance}}

\section{Discussion}\label{sec:conclusion}

Our main contribution is the systematization of the rather fragmented 
body of literature related to decentralization into 
a unified framework, under which 
the decentralization of any distributed ledger can be analyzed.
Table~\ref{tab:summary} summarizes our 
methodology, \ie the layers that comprise a ledger and the 
relevant resources and parties that guarantee its core 
properties. Removing single points of failure via
a diverse distribution of resources across independent parties is
critical in guaranteeing security, privacy, and stability
and can also have legal implications 
\ifsubmission
(cf. Appendix~\ref{sec:regulation}). 
\else
(cf. Definition~\ref{def:mdt}). 
\fi
Our work also opens various research threads.

%
First, even though we do not treat off-chain ``Layer 2'' 
protocols~\cite{EPRINT:JKLT19} as a distinct stratum, our 
methodology can be readily applied to a
combination of Layer 1 (main chain) and Layer 2 
protocols. For example, one could apply our layering methodology 
on a combination of Bitcoin and the Lightning 
Network~\cite{poon2016bitcoin} and assess the decentralization 
of the combined system as a payment network. Similarly,
our methodology can assess blockchain-based ``decentralized 
applications'' (DApps), 
\eg DeFi systems.\footnote{In such cases, 
smart contract development would naturally fall in the 
software layer.} Investigating the 
decentralization of Layer 2 protools or DApps 
via our methodology
is thus a promising research thread.

Second, exploring the relationship between 
decentralization and fault tolerance, as well as the 
settings where decentralization is beneficial and those where it 
is not, is another interesting topic of future research.\footnote{Appendix
\ref{sec:bft} expands on the relationship between fault tolerant systems and
decentralization.} 


Third, we offer a framework for analyzing blockchain 
decentralization, but not specific quantitative metrics.
A compelling direction for future work is
systematically identifying the right metrics for 
each layer to capture all relevant aspects of blockchain 
decentralization.\footnote{Appendix \ref{sec:metrics} offers an overview of 
metrics that have been used so far in this setting.} 
Building on this, a natural end-goal of all questions 
posed above is producing a quantitative 
blockchain decentralization index. Historically, it has been 
observed that increasing decentralization on one axis coincides 
with, or even results in, centralization on 
another~\cite{schneider2019decentralization}.
Future work should further explore the dynamics between all 
layers, resolve possibly inescapable trade-offs,
determine the importance (weight) that should be assigned to each
layer and each metric, and efficiently combine them into an 
index.



\iflipics
\bibliographystyle{plainurl}
\else

\ifieee
    \bibliographystyle{format/IEEEtran}
\else
    \ifccs
        \bibliographystyle{format/ACM-Reference-Format}
    \else
        \iflncs
            \bibliographystyle{format/splncs04}
        \else
            \bibliographystyle{alpha}
        \fi
    \fi
\fi

\fi
\bibliography{bibliography/additional,bibliography/cryptobib/abbrev0,bibliography/cryptobib/crypto_crossref}

\ifieee
\appendices
\else
\appendix
\fi

\ifsubmission


%

\fi

\ifshort

\section{Brief Evaluations per Layer}\label{sec:evaluations}

In this section we provide brief evaluations of various systems for each
subcategory of the layers covered in Sections~\ref{sec:hardware}
-~\ref{sec:geography}. In doing so, we identify various questions that require
further research across two broad axes. First, from a measurement perspective,
many systems and dimensions lack pertinent data or, to make matters
more interesting, it is unclear how to even conduct robust measurements for the
data under question. Second, from a design perspective, a relevant thread of
research would focus on enabling or incentivizing protocol designers to
implement accurate data collection mechanisms as a part of the systems
themselves.

\subsection{Hardware: Physical Hardware}
\ifshort
\else
\subsubsection*{Evaluation}
\fi
Centralization around specialized hardware has been
documented~\cite{ekblaw2016bitcoin}, although no academic research could be
found on mining hardware usage in real-world systems. Interestingly, it is
unclear how to even measure the usage of hardware equipment in PoW mining via public data, as
well as how to develop PoW algorithms that promote diversity, thus future
research could aim at answering these questions.
Nonetheless, there exist some reports,
though they often present conflicting assessments. In Bitcoin,
between $2017-2019$ a single mining hardware provider accounted for either
$65-75$\%~\cite{bendiksen2019bitcoin} or $46-58$\%~\cite{tokeninsight-2} of the
network's hashrate, with $98$\% of the market controlled by $4$ firms.


\subsection{Hardware: Virtual Hardware}
\ifshort
\else
\subsubsection*{Evaluation}
\fi
A comprehensive evaluation of centralization in terms of hardware hosting, and 
how to incentivize hosting diversity is a promising thread of future research.
Here, we consider two examples of highly-valued PoS
systems, Solana and Avalanche.\footnote{\href{https://solana.com}{Solana} and
\href{https://www.avax.network/}{Avalanche} are \#9 and \#14 respectively \wrt
market capitalization. [\href{https://coinmarketcap.com}{CoinMarketCap}; August
2022]} 
Solana's validators predominantly operate cloud-based nodes; of the $1873$
nodes, more than half are hosted in two services, with more than $50$\% and
more than $66$\% of participating stake hosted by $3$ and $5$ providers
respectively.\footnote{\href{https://www.validators.app/data-centers?locale=en\&network=mainnet\&sort\_by=asn}{validators.app}.
[August $2022$]}
Avalanche observes similar concentration issues; $731$ out of $1254$ validators,
who control $71.84$\% of all stake, are hosted by a single
company.\footnote{Data obtained from
\href{https://avascan.info/stats/staking}{avascan.info}. [August $2022$]} 


\subsection{Software: Protocol Participation}
\ifshort
\else
\subsubsection*{Evaluation}
\fi
The literature is lacking formal analyses on the usage of full node
ledger software, so a rigorous evaluation of the dynamics in software development
and usage could highlight various centralization tendencies.
%
Various community and commercial projects do keep track of
statistics though. In most systems, a single client software is
predominantly used by the participating nodes in the network. In Bitcoin
$99$\% use Bitcoin Core (aka Satoshi), in Ethereum $78$\% use geth, in
Litecoin $95$\% use LitecoinCore, while systems like Zcash are completely
centralized with all nodes using one software (MagicBean); a notable exception
is Bitcoin Cash, where usage is split between BCH Unlimited ($33$\%), Bitcoin
Cash Node ($51$\%), and Bitcoin ABC ($12$\%).\footnote{Sources:
\href{https://blockchair.com}{blockchair},
\href{https://ethernodes.org/}{ethernodes} [August 2022]} 


Some projects are actively managed by a wide network of
developers, \eg more than $200$ contribute to Ethereum~\cite{devtrends}, while others are particularly
centralized. As of $2018$, $7$\% of all Bitcoin Core files were written by the
same person, while $30$\% of all files had a single author. In Ethereum,
these figures rise to $20$\% and $55$\% respectively~\cite{FCW:AzoMalMei18}.
Comments observe similar centralization patterns, with $8$ ($0.3$\%) and $18$
($0.6$\%) people contributing half of all comments in Bitcoin and Ethereum
respectively~\cite{FCW:AzoMalMei18}.

\subsection{Software: Asset Management}

\subsection{Network: Topology}
\ifshort
\else
\subsubsection*{Evaluation}
\fi
Bitcoin is notoriously vigilant in hiding its network topology~\cite{FC:DBPLPMB19,franzoni2022atom}. 
Various works analyze it by inferring
a node's neighborhood~\cite{CCS:BirKhoPus14}, timing
analysis~\cite{neudecker2016timing}, or conflicting
transaction propagation~\cite{FC:DBPLPMB19}. In $2014$, it was found that more than half
of Bitcoin nodes resided in $40$ autonomous systems (ASs), with $30$\% in just
$10$ ASs~\cite{DBLP:conf/ant/FeldSW14}. In $2017$, Bitcoin's and
Ethereum's P2P networks observed similar sizes ($3390$ nodes for Bitcoin, $4302$
for Ethereum). Bitcoin offered lower latency and higher
bandwidth, with nodes being closer geographically and $56$\% of
them hosted on dedicated hosting services (vs. $28$\% for
Ethereum)~\cite{FC:GBEvS18}. In addition,
$68$\% of the mining power was hosted on $10$ transit networks, while $3$ transit
networks saw more than $60$\% of all connections~\cite{SP:ApoZohVan17}.
In $2019$, Ethereum's network presented a large degree of
centralization around clusters, forming a ``small world
network''~\cite{DBLP:conf/iscc/0003SWTZY19} with $10$ cloud hosting
providers accounting for $57$\% of all nodes and one hosting
almost a quarter~\cite{chainstack}. This was reaffirmed in $2020$, as Ethereum
messages could be sent to most nodes within $6$ hops~\cite{wang2021ethna}.
In $2020$, Monero's topology also observed a high level
of centralization, as $13.2$\% of nodes maintained $82.86$\% of all
connections~\cite{FC:CYDLV20}. No analysis of PoS systems'
networks could be found; given their non-reliance on specialized hardware and ease of
relocation, a PoS-PoW comparison would be of interest.

Bitcoin, as the first blockchain system, has also seen multiple eclipse attacks
and defenses~\cite{SP:TCMVK20,USENIX:HKZG15,USENIX:TraSheKan21}. Some works
attempt to increase the number of connections
without reaching prohibitive levels of bandwidth
usage~\cite{CCS:NMWFB19}.
Ethereum was also found vulnerable to eclipse attacks that do not require
monopolizing a node's connections, but relied on message
propagation~\cite{wust2016ethereum}.

%

\subsection{Network: Node Bootstrapping and Peer Discovery}
\ifshort
\else
\subsubsection*{Evaluation}
\fi
Bitcoin Core defines $8$ outgoing connections, selected randomly from a known list of identities, and up to
$125$ incoming~\cite{survey2021}. When (re)joining the network, a node
attempts to connect to previously-known identities and, if unsuccessful,
employs a (hardcoded) list of DNS seeds. Other systems, like Ethereum and
Cardano, employ more complex, DHT-based mechanisms~\cite{maymounkov2002kademlia} that require further 
analysis. Cardano is also an interesting
implementation, as it assumes two node types:
\begin{inparaenum}[(a)]
    \item core nodes that participate in consensus, and
    \item relays that intermediate between core and edge nodes (e.g., wallets);
\end{inparaenum}
in the default configuration relays  are operated by only a small committee~\cite{survey2021}.

\subsection{Consensus}
\ifshort
\else
\subsubsection*{Evaluation}
\fi
In (game) theory, Bitcoin's resistance to centralization has been both
supported \cite{kroll2013economics,DBLP:conf/aft/KiayiasS21} and
\ifshort
refuted~\cite{DBLP:journals/cacm/EyalS18},
\else
refuted~\cite{FC:EyaSir14,DBLP:journals/access/LiuKXJW19},
\fi
depending on the economic model assumed for the participants' utilities. In practice,
mining pools have been observed as early as
$2013$~\cite{DBLP:journals/ieeesp/GervaisKCC14}. Between $2016-2020$,
pools created $98.6$\% of Bitcoin
blocks~\cite{DBLP:journals/corr/abs-2002-02082}, with $5$ pools consistently
contributing between $65-85$\% of the eventual blocks and $25$
controlling more than $94$\% of all hashing
\ifshort
power~\cite{measurement2020}.
\else
power~\cite{measurement2020,DBLP:conf/icde/LinLZC21}. 
\fi
Centralization has also been
observed within mining pools. Between $2017-2018$, no entity
controlled more than $21$\%~\cite{FC:GBEvS18} of hashing power, but three pools controlled 
a majority; within these pools, a few participants
($\leq 20$) received over $50$\% of
rewards~\cite{DBLP:journals/corr/abs-1905-05999}. Miners often
participate in multiple pools at the same time, a behavior also observed in
Ethereum~\cite{DBLP:conf/infocom/ZengCCZGXM21}. Although
centralization around pools is high in (PoW-based) Ethereum (in $2019$, $3$
pools controlled a majority of mining
power~\cite{lin2021measuring}), power within the pools is
spread across hundreds of
addresses~\cite{DBLP:conf/infocom/ZengCCZGXM21}, albeit some possibly
owned by the same parties.

\subsection{Cryptocurrency Economics: (Initial) Token Distribution}
\ifshort
\else
\subsubsection*{Evaluation}
\fi
PoS systems like Cardano, NEO, and Algorand tried to reduce early-stage risks
via a two-phase launch. At first, the ledger was  
controlled by either the core development company or
foundation or a committee numbering a small number of
entities. After token ownership was sufficiently distributed,
participation opened widely to all stakeholders.
Beyond the obvious issues in maintaining a
permissioned database, the first phase typically takes years to conclude.
Early users often tend to either not
participate or transfer their tokens to the few exchanges
that support these new tokens~\cite{victor2021taxonomy}. Therefore, an interesting
question is the relationship of the delay between launch and full
decentralization and the diversity
of early investors.
%

\subsection{Cryptocurrency Economics: Token Ownership}

\subsection{Cryptocurrency Economics: Secondary Markets}

\subsection{Client API}
\ifshort
\else
\subsubsection*{Evaluation}
\fi
In Bitcoin, most wallets are either SPV or
explorer-based~\cite{karantias2020sok}. In the first case, the wallet obtains
the chain's headers and, to verify that a transaction is published,
requests a proof from full nodes. Although SPV does mitigate safety attacks, it
also hurts the user's privacy, as their transaction information is leaked to
the full node operators. Explorer-based wallets instead rely entirely on a 
single explorer service and its full nodes, which are trusted completely.
In $2018$, $5-10$\% of all Ethereum nodes reportedly relied on a
centralized blockchain API service, Infura~\cite{infura-usage}. This reliance
continued throughout the years. In $2020$, a service outage demonstrated in
practice the hazards of such centralization~\cite{infura-outage}. In $2022$, a
misconfiguration on Infura's part resulted in wallets (and, thus, user funds)
being inaccessible~\cite{infura-bug}.  In terms of applications, OpenSea is the
leading hosting service for NFTs. As of $2021$, it reportedly handled $98$\% of
all NFT volume~\cite{opensea-usage}, charging a $2.5$\% commission on all
sales. As expected, an OpenSea outage in $2022$ also resulted in the NFT
market being practically unusable~\cite{opensea-outage}.

\subsection{Governance: Improvements and Conflict Resolution}
\ifshort
\else
\subsubsection*{Evaluation}
\fi
Most systems employ an \emph{Improvement Proposal} mechanism,
where proposals are posed as issues in Github, a (centralized) system that is extensively used for software development. If a change gathers enough support,
it is incorporated in the codebase. To voice approval for proposals, miners
often include encoded messages in blocks.
From early on, proposals in Bitcoin and Ethereum have been made by a
handful of developers~\cite{DBLP:journals/ieeesp/GervaisKCC14,FCW:AzoMalMei18}.
In the discussion phase, many people participate but again only a
few actors contribute most comments, while in cases like
Bitcoin the groups of developers and commenters largely overlap.

\subsection{Governance: Development Funding}
\ifshort
\else
\subsubsection*{Evaluation}
\fi
Most existing blockchain systems follow the first approach, \ie not making
funding provisions. In many cases,
funding is channeled through a few foundations and
companies.\footnote{The first usually take the name of the token, \eg the
\{Bitcoin, Ethereum, Cardano\} Foundations. Examples of the second are the ASIC
companies discussed in Section~\ref{subsec:physical-hardware} or software 
companies
like Blockstream (Bitcoin), Consensys (Ethereum), Input Output (Cardano), \etc}
Treasuries are present in some ledgers, like Decred, Cardano, and Dash.
Despite their potential though, widespread funding has yet to be 
demonstrated for most systems.\footnote{Decred's treasury holds
\$$23.8$M, and has allocated $\$250$K over the past year. Cardano's
treasury holds approx. \$$500$M and has distributed \$$17.2$M across $939$
projects. Dash, one of the first systems to set a treasury, allocated
\$$500,000$ over $2018$, but it appears non-functional as of
$2022$. 
[\href{https://dcrdata.decred.org/treasury}{dcrdata.decred.org},
\href{https://bit.ly/FundedProjectsReporting}{cardano.ideascale.com},
\href{https://dashvotetracker.com}{dashvotetracker}; August $2022$]}

\subsection{Geography}

\section{Measuring decentralization}\label{sec:metrics}
Our work offers a framework for analyzing blockchain decentralization, but
not specific metrics to quantitatively measure it. For example, a metric could 
assign a
single number to reflect how close a system is to a single
point of failure, given a distribution of resources over a set of relevant
parties. Here, we briefly review some metrics,
at a high level, and leave for future work the exploration of alternatives and the computations over
real-world data.
A first option is Shannon entropy~\cite{shannon2001mathematical}. Briefly, a random
variable's entropy measures the uncertainty of its possible
outcomes. In our setting, the more bits of entropy in the resource
distribution, the more diverse it is, thus the more decentralized the measured
component is. Min-entropy, \ie the smallest of the R{\'e}nyi family of
entropies~\cite{renyi1961measures} can be also used instead since it also offers
a lower bound. 
An alternative is the Gini coefficient~\cite{ginicryptos}. Gini
expresses the percentage of space between the $45^o$ line and the curve
that plots the cumulative wealth $y$ owned by the bottom $x$ of the
population. Intuitively, a Gini value of $0$ implies perfect equality, where
each person owns the same amount of resources, while $1$ reveals extreme
inequality.\ifshort~\else\footnote{Although Gini is used by organizations like the
OECD~\cite{oecdGini} and the World Bank~\cite{world-bank-gini}, it has major
shortcomings. First, \emph{what} the metric is applied on may skew the
narrative. E.g., the Gini coefficient of real-world economies \wrt to
income offers a much more appealing image compared to wealth ownership, as
income differences persist and accumulate over the course of life. Second,
Gini's single value obscures qualitative differences between economies at the
same ``level''. For instance, consider the following two wealth
distributions~\cite{vitalik-gini}:
i) one person holds $50$\% of total wealth, while all else share the rest;
ii) half the population shares equally all wealth, while the other half has
none.
Although these economies are vastly different, their Gini coefficient is the
same (approx. $0.5$).}\fi
Alternative metrics could
also help evaluate different aspects of decentralization. Examples from traditional economics 
are the Theil~\cite{theil1967economics},
Atkinson~\cite{atkinson1970measurement}, and Herfindahl-Hirschman~\cite{rhoades1993herfindahl} indices.
Drawing from the blockchain space, an often-used metric is the Nakamoto
coefficient~\cite{nakamoto-coefficient}, which measures the minimum number of
parties that control a majority of resources. Nonetheless, a systematic
comparison of all alternatives is an interesting question for future research.
%

\section{Fault Tolerance and 
Decentralization}\label{sec:bft}
Decentralization disperses control across a large set of parties. This is
seemingly beneficial for Byzantine Fault Tolerant (BFT) systems. On the other hand, it may be
counterproductive for other notions of faults.
Specifically, the goal of BFT systems is to sustain corruptions of some
(bounded) number of participants. Therefore, avoiding single points of failure
and distributing the system's operation is particularly useful in this context.
The more decentralized a BFT system is, the more parties an adversary needs to
corrupt.
Non-BFT systems, which are \eg crash fault tolerant, may not be able to sustain the
corruption of any participant. In other words, even if a single participant
behaves in a Byzantine manner, the system's properties cannot be
guaranteed. Thus, the more decentralized a non-BFT system is, the larger its
attack surface. Therefore, its security relies on the security of the weakest
participant. For larger numbers of participants, \ie if the system is more
decentralized, the likelihood that an adversary can corrupt any one
participant typically increases, since participants often do not have the same
level of security.
Therefore, exploring the relationship between decentralization and 
fault tolerance, as well as the settings where decentralization is beneficial
and those where it is not, is another interesting topic of future research.

\fi

\end{document}